\documentclass[showpacs,preprintnumbers,amsmath,amssymb]{revtex4}
\usepackage{graphicx,color}
\usepackage{bm}
\usepackage{color}

\begin{document}

\title{Non-axisymmetric baby-skyrmion branes}
\author{T\'{e}rence Delsate$^a$}
\email{terence.delsate(at)umons.ac.be}
\author{Masahiro Hayasaka$^b$}
\email{dxdt2@i.gmobb.jp}
\author{Nobuyuki Sawado$^b$}
\email{sawado(at)ph.noda.tus.ac.jp}
\affiliation{$^a$Theoretical and Mathematical Physics Dpt.,
Universit\'{e} de Mons -\\ UMons, 20, Place du Parc, 7000 Mons - Belgium \\
$^b$Department of Physics, Faculty of Science and Technology,
Tokyo University of Science, Noda, Chiba 278-8510, Japan}

\date{\today}

\begin{abstract}
We investigate the existence of non axisymmetric solutions in the $6$-dimensional baby-Skyrme brane model. The brane is described by a localized solution to the baby-Skyrme model extending in the extra dimensions. Such non symmetric branes have already been constructed in the original $2+1$-dimensional baby-Skyrme model in flat space. We generalize this result to the case of gravitating baby-Skyrme and in the context of extradimensions. 
These non-trivial deformation from the axisymmetric shape appear for higher values of the topological charge, so we consider the cases of $B=3,4$, where $B$ is the topological charge. 
We solve the coupled system of the Einstein and baby-Skyrme equations 
by successive over relaxation method. We argue that the result may be a possible resolution for the fermion mass hierarchy puzzle.
%
%
\end{abstract}

\maketitle

\section{Introduction}
Theories with extradimensions are expected to solve the hierarchy 
problem and the cosmological constant problem \cite{ccp}. The great challenge with extradimensions is to conciliate the extradimensions with the fact that we observe only four dimensions. 
The typical way is to assume that we live on branes \cite{dvali}. The branes are $4$-dimensional subspace of the entire extradimensional space-time where the matter fields are trapped. 
One possible extradimensional geometry was proposed by Randall-Sundram (RS), where the extradimensions is warped. The RS brane model is formulated in five space-time 
dimensions~\cite{Randall:1999ee,Randall:1999vf} and shows that the exponential 
warp factor in the metric can generate a large hierarchy of scales. However, such models require a fine tuning between the parameters of the model. 
Note that in the RS model, the brane is infinitely thin. Such a fine tuning is usually not needed if thick branes are considered. 
For instance, the brane theories in six dimensions in the abelian strings 
show a very distinct feature towards the fine-tuning and negative tension brane problems
~\cite{Cohen:1999ia,Gregory:1999gv,Gherghetta:2000qi,
Giovannini:2001hh,Peter:2003zg}. 
Similar compactification was achieved for magnetic monopoles, both for positive and negative cosmological constant~\cite{Roessl:2002rv}. 

The Skyrme model~\cite{Skyrme:1961vq} is originally a model for the nucleons after a suitable
quantization scheme~\cite{Adkins:1983ya}.
There is a low-dimensional mimic of the model called baby-Skyrme model. 
The baby-Skyrme model possesses soliton solutions called the baby-skyrmions in two dimensional 
space~\cite{Leese:1989gi,Piette:1994ug,Kudryavtsev:1996er,Kudryavtsev:1997nw}. 
The warped compactification of the two dimensional extra space by such baby-skyrmions was already studied~\cite{Kodama:2008xm} for negative bulk cosmological constant, based on the assumption that the cosmological constant inside the three branes is tentatively set to be zero. 
Addressing the non-zero cosmological constant inside the branes has been considered for case of the strings~\cite{Brihaye:2006pi} and the monopoles~\cite{Brihaye:2006cs}. Along these directions, 
we also have studied the baby-skyrmion brane with both positive brane cosmological constant and a bulk cosmological 
constant~\cite{Brihaye:2010nf}. 

The localization of particles such as fermions~\cite{Kodama:2008xm,Brihaye:2010nf,Delsate:2011aa} 
and a gauge field~\cite{Delsate:2011aa} on these branes have also been studied. 
According to the Index theorem a nonzero topological charge implies a
zero modes of the Dirac operator \cite{Atiyah:1980jh}. 
The zero crossing modes are found to be the localized fermions on the brane. 
It follows that the generation of the fermions emerges as an effect of the topological 
charge of the skyrmions with a special quantum number assigned by $K_3$. 
There are different profiles of the zero crossing behavior for different $K_3$, 
and in our opinion, this might give an origin of the finite mass generation fermions.
In this paper we begin our analysis for the topological charge $B=3$ because it is the number of particle generations in the fermions.
As the straightforward extension, the case of $B=4$ is also examined. 

In our previous studies, we have found an essential requirement for the geometry of the brane solutions:
the brane should slightly deform from the axisymmetric shape in order to explain the small mass 
difference between the first two generations of quarks (such as $M_s-M_u \sim 10^2$ MeV, which is much smaller 
than the difference of the first and third: $M_t-M_u \sim 10^6$ MeV). In \cite{Kodama:2008xm}, we 
introduced an effect of the deformation for the matter fields by hand and 
found that about for a slight deformation of the skyrmion the result is in good agreement with the experiment. 
The drawbacks of the analysis are:  (i) we need to introduce an arbitrary deformation parameter, (ii) we ignore effect
of the deformation for the gravity side.  
Our main concern of this paper is thus to find brane solutions with spontaneously broken axisymmetry. 
The solutions with lower symmetry, such as $\mathbb{Z}_2$-symmetry without gravity were already found using numerical 
simulation~\cite{Hen:2007in} where an energy minimization scheme 
called the {\it  simulated annealing method}~(SA)~\cite{Hale:2000fk} was used. For the analysis of the brane solutions, 
the coupled system of matter field equations and Einstein equations has to be treated and 
the SA scheme is not suitable for the purpose. Therefore, as a first step, we solve baby-Skyrme field equations without 
gravity by employing the standard {\it successive over-relaxation method}~ (SOR) and compare the obtained results by the SA as a validation of our code.  

For the brane solutions, we use an ansatz for the line element which is a function 
of both the two extra coordinates $(r,\theta)$, inspired by the Lewis-Papapetrou ansatz. 
The resulting coupled partial differential equations are numerically solved in terms of 
the SOR~\cite{relaxation}. 

This paper is organized as follows. 
In Sec.II, we start with a brief introduction to the baby-Skyrme model and show the results using SA. We
also present the Euler-Lagrange equations of the non symmetric solutions. 
The corresponding brane model is extensively discussed in Sec.III. In Sec.IV we present the numerical solutions and give 
a brief summary of our results in Sec.V.

\section{The Baby-Skyrme model and the deformed solutions}
The lagrangian density of the baby-Skyrme model is given by
\begin{eqnarray}
{\cal L}_{\text{baby}} =\frac{\kappa_{2}}{2}(\partial_{M}\bm{n})\cdot(\partial^{M}\bm{n})-\frac{\kappa_{4}}{4}(\partial_{M}\bm{n}\times\partial_{N}\bm{n})^{2}- \kappa_{0}V(\bm{n}),
\label{eq:action}
\end{eqnarray}
where $\bm{n}=(n_1,n_2,n_3)$ denotes a triplet of real scalar fields with the constraint $\bm{n}\cdot\bm{n}=1$. The constants
$\kappa_{2,4,0}$ are parameter with dimension [$M$],~[$M^{-1}$],~[$M^{3}$] respectively, where $M$ is the dimension of mass.
The first term in (\ref{eq:action}) is nothing but the non-linear $\sigma$ model while the second term is inspired by the quartic term of the $3+1$ Skyrme model. The potential $V$ is necessary for the stability of the solutions in the sense of Derrick's argument. 
While the first two terms are invariant under a global $O(3)$ symmetry, $\bm{n}\to Q\bm{n},Q\in O(3)$, the potential breaks it. 
Several choices of potentials have been investigated. 
Two $O(2)$-symmetric potentials have been intensively studied, namely the old-type $V(n_3)=(1+n_3)$ \cite{Leese:1989gi}
and the new-type potential $V(n_3)=(1-n_3^2)$~\cite{Kudryavtsev:1997nw}. 
Potentials with no $O(2)$ symmetry have been studied in \cite{Ward:2003xv,Jaykka:2011ic}. 
In \cite{Hen:2007in}, the authors generalized the old-type 
potential to $V(n_3)=(1+n_3)^s$ and were able to build solutions that deviate from the axial symmetry.
The model admits an analytical solution (holomorphic solution) for a single
soliton for $s=4$. The force between two holomorphic solutions is always repulsive so that multi-soliton solutions tend to deviate from axial symmetry~\cite{Leese:1989gi}. 

The topological charge of the baby-skyrmion is defined as
\begin{eqnarray}
B=\frac{1}{4\pi}\int d^2x~\bm{n}\cdot (\partial_1 \bm{n}\times\partial_2\bm{n})
\label{topologicalcharge}
\end{eqnarray}
in $2+1$ flat spacetime. 
The numerical analysis in \cite{Hen:2007in} has been done by minimizing the following energy functional 
\begin{eqnarray}
E=\int^{\pi}_{-\pi}d\theta \int^\infty_0 \rho d\rho \Bigl\{
\frac{\kappa_2}{2}\Bigl((\partial_\rho\bm{n})^2+\frac{1}{\rho^2}(\partial_\theta\bm{n})^2\Bigr)
+\frac{\kappa_4}{2}\frac{1}{\rho^2}(\partial_\rho\bm{n}\times\partial_\theta\bm{n})^2+\kappa_0V(\bm{n})
\Bigr\}
\label{senergy}
\end{eqnarray}
in the polar coordinate $(\rho,\theta)$. The method is a kind of the Monte Carlo method in which one generates 
random numbers and properly change the value of the fields $\bm{n}$ by the numbers so as to fall down the energy.
The more sophisticated method may be applied to the problem.  
The {\it simulated annealing method}~(SA)~\cite{Hale:2000fk} is the application of the Metropolis algorithm 
which can successfully avoids the unwelcome saddle points.

\begin{figure}[t]
\includegraphics[clip,width=50mm]{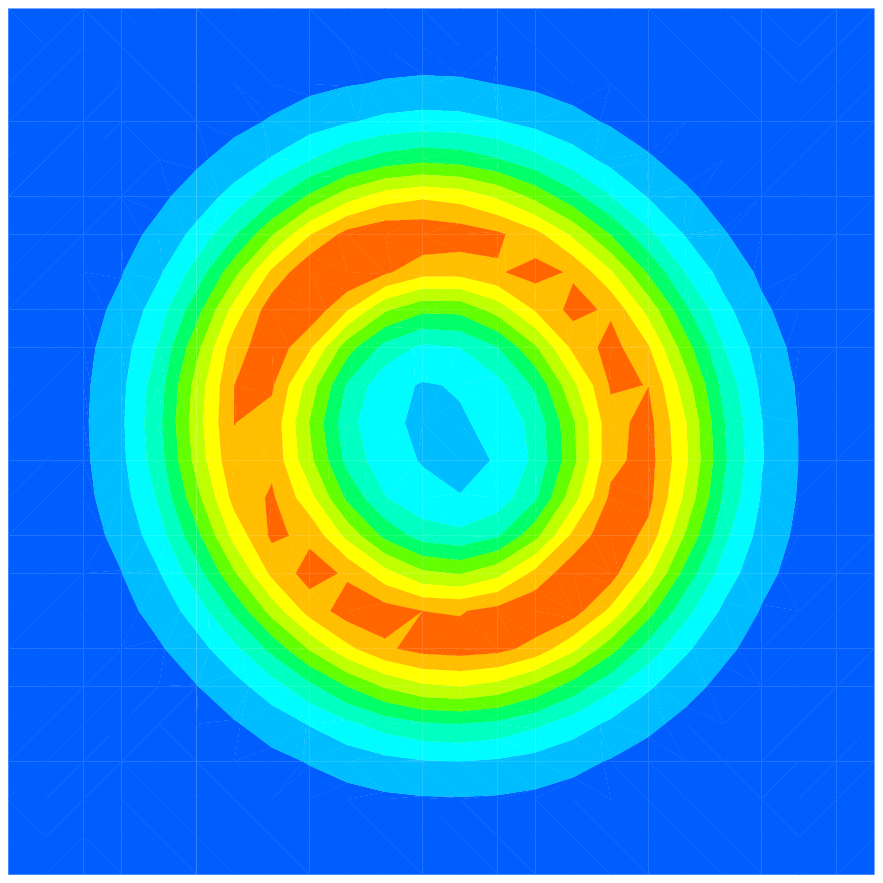}\hspace{-1cm}
\includegraphics[clip,width=50mm]{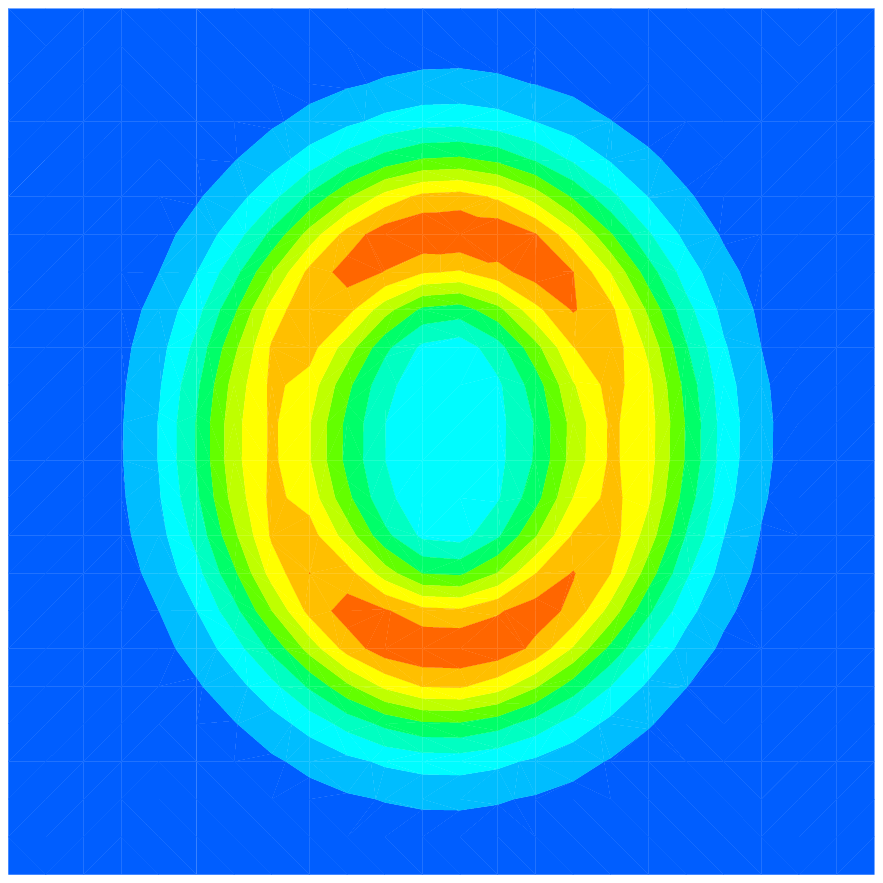}\hspace{-1cm}
\includegraphics[clip,width=50mm]{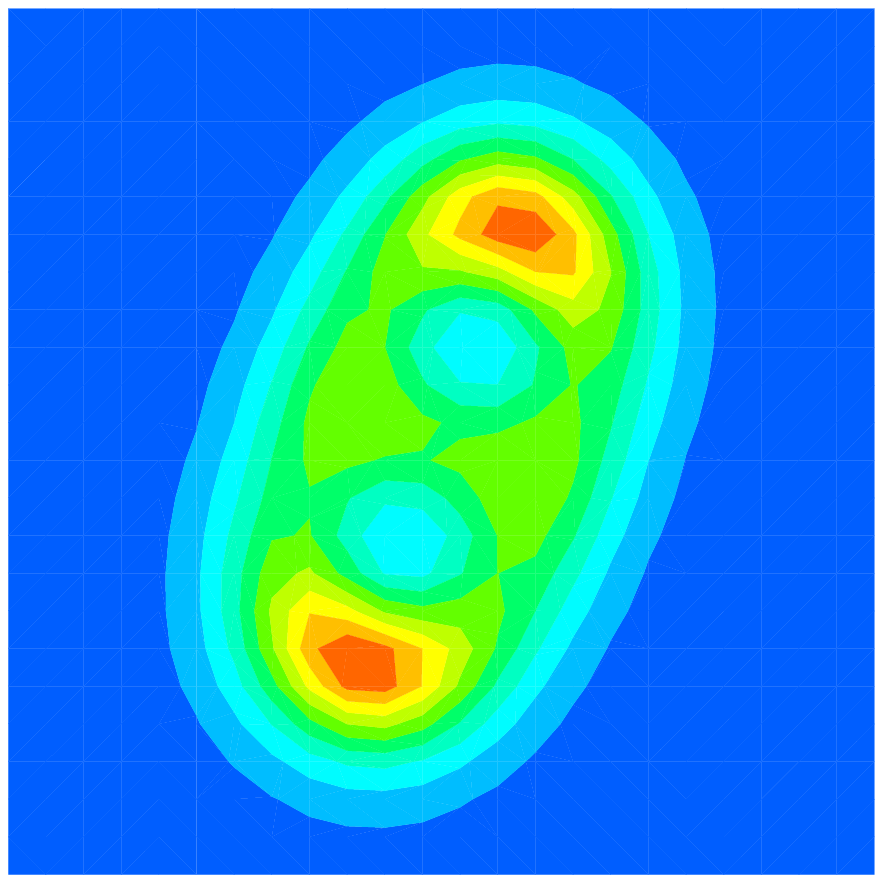}\hspace{-1cm}
\includegraphics[clip,width=50mm]{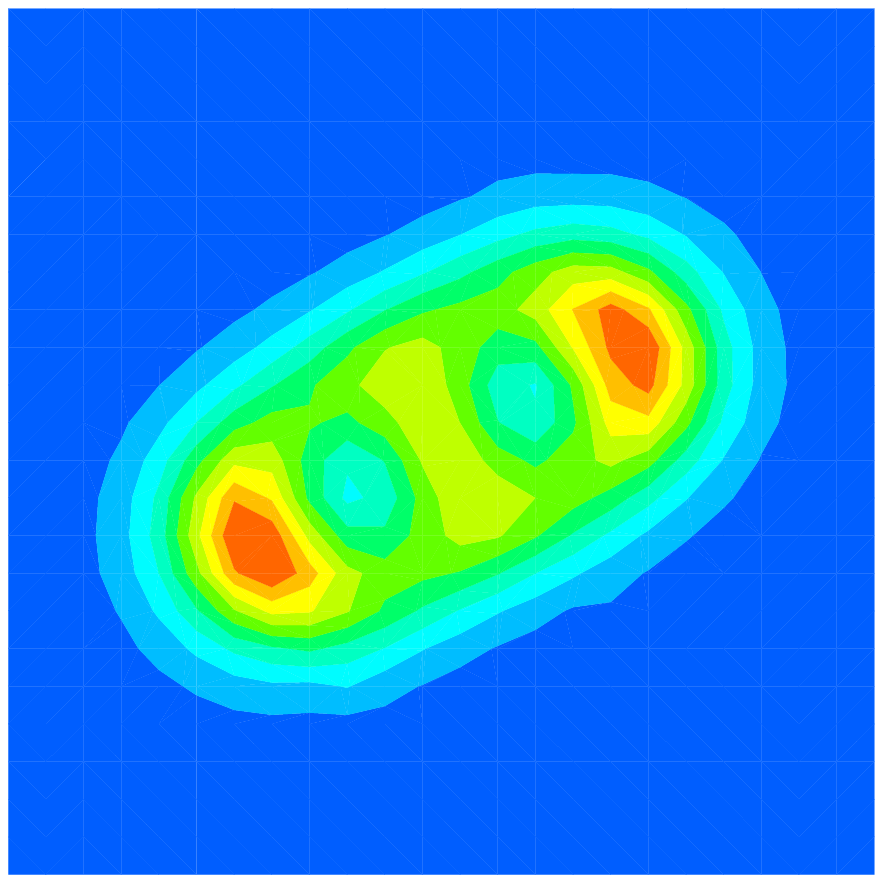}
\caption{\label{fig:annealing_e}Contour plot of the energy densities of the $B=3$ baby-skyrmion 
for several strength of the parameters $\mu:=\kappa_0\kappa_2/\kappa_4^2$. From the left $\mu=0.04,0.06,0.12,0.64$ with $s=1$.}
\end{figure}

\begin{figure}[t]
\includegraphics[clip,width=50mm]{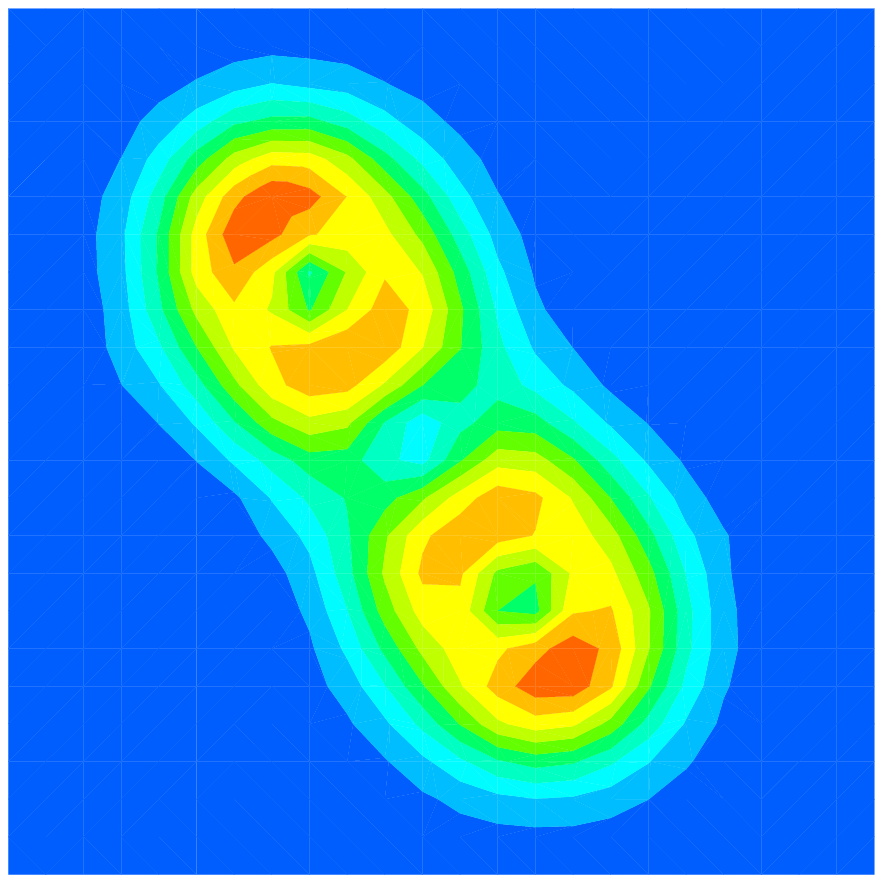}\hspace{-1cm}
\includegraphics[clip,width=50mm]{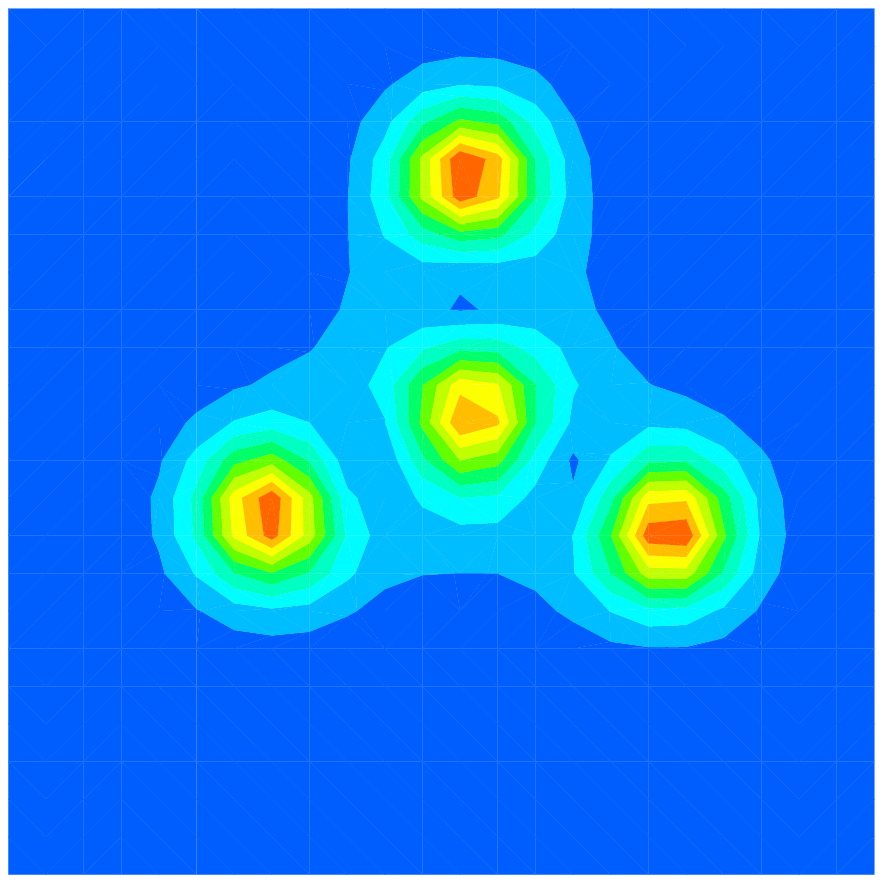}
\caption{\label{fig:annealing_e4}Contour plot of the energy densities of the $B=4$ baby-skyrmion 
for the parameter $\mu=0.64,s=1$ (left) and $\mu=0.32,s=2$.}
\end{figure}

The solutions for $B\geqq 2$ exhibit a non-axisymmetric shape 
for certain parameter ranges of the coupling constant $\kappa_0$ or of the value of $s$. 
We confirm the results of \cite{Hen:2007in} by minimizing the energy per topological charge in terms of the SA.
In Fig.\ref{fig:annealing_e}, we plot the typical results of the energy densities of the skyrmion with varying $\kappa_0$.
As was found in \cite{Hen:2007in}, at a critical value of $\kappa_0$, breaking of the rotational symmetry occurs and 
only the $\mathbb{Z}_2$-symmetry survives for $B=3$. 
More interestingly, the case of $B=4$, the $\mathbb{Z}_2$ remains for $s=1$, 
whereas the $s=2$ the solution exhibits a $\mathbb{Z}_3$-symmetry (see Fig.\ref{fig:annealing_e4}).
We restrict our numerical analysis for the case of $s=1$ except for some results of $B=4$.

In this section, we solve the Euler-Lagrange equations derived from the 
action (\ref{eq:action}) and try to reproduce the results of the SA. 
Generally speaking,  both results should coincide. However, we need to introduce additional boundary conditions in order to 
solve the equations using a SOR algorithm which possibly strongly affect the properties of the solutions. 

We use the following ansatz for the scalar triplet $\bm{n}$:
\begin{equation}
	\bm{n} = (\sin F(\rho,\theta)\cos\Theta(\rho,\theta), \sin F(\rho,\theta)\sin\Theta(\rho,\theta), 
	\cos F(\rho,\theta)),
	\label{eq:hedgehog-ansatz}
\end{equation}
where $\Theta(\rho,\theta):=\Theta_0(\rho,\theta)+n\theta$ where $\Theta_0(\rho,\theta)$: $S^1\mapsto S^1$ 
is homotopic to the constant map. 

The resulting Euler-Lagrange equations are of the form
\begin{eqnarray}
			&&r^{2}\partial^{2}_{r}F
			+\partial^{2}_{\theta}F
			+r\partial_{r}F
			-\frac{1}{2}\sin2F\bigl(r^{2}\left(\partial_{r}\Theta\right)^{2}
			+\left(\partial_{\theta}\Theta\right)^{2}\bigr) \nonumber \\
			&&+\frac{1}{r^{2}}\sin^{2}F\biggl\{
			-\frac{1}{r}
			\biggl(\partial_{r}F\partial_{\theta}\Theta-\partial_{\theta}F\partial_{r}\Theta\biggr)\partial_{\theta}\Theta\biggr.
			+\partial^{2}_{r}F\left(\partial_{\theta}\Theta\right)^{2}
			+\partial^{2}_{\theta}F\left(\partial_{r}\Theta\right)^{2}
			+\biggl(\partial_{r}F\partial_{\theta}\Theta+\partial_{\theta}F\partial_{r}\Theta\biggr)\partial_{r}\partial_{\theta}\Theta
			\nonumber \\
			&&-\biggl(\partial^{2}_{r}\Theta\partial_{\theta}\Theta\partial_{\theta}F
			+\partial^{2}_{\theta}\Theta\partial_{r}\Theta\partial_{r}F
			+2\partial_{r}\partial_{\theta}F\partial_{\theta}\Theta\partial_{r}\Theta\biggr)
			\biggr\}
			+\frac{1}{2r^2}\sin2F\biggl(\partial_{r}F\partial_{\theta}\Theta-\partial_{\theta}F\partial_{r}\Theta\biggr)^{2}
			-\mu\frac{\partial V}{\partial F}=0\,, 
\end{eqnarray}
\begin{eqnarray}
&&			r^{2}\partial^{2}_{r}\Theta
			+\partial^{2}_{\theta}\Theta
			+r\partial_{r}\Theta
			+\cot F\bigl(r^{2}\partial_{r}F\partial_{r}\Theta
			+\partial_{\theta}F\partial_{\theta}\Theta\bigr)
			\nonumber \\
&&			-\frac{1}{r^{2}}\biggl\{
			-\frac{1}{r}
			\biggl(\partial_{r}F\partial_{\theta}\Theta-\partial_{\theta}F\partial_{r}\Theta\biggr)\partial_{\theta}F\biggr.
			-\partial^{2}_{r}\Theta\left(\partial_{\theta}F\right)^{2}
			-\partial^{2}_{\theta}\Theta\left(\partial_{r}F\right)^{2}
			-\biggl(\partial_{r}F\partial_{\theta}\Theta+\partial_{\theta}F\partial_{r}\Theta\biggr)\partial_{r}\partial_{\theta}F
			\nonumber \\
&&			\biggl.
			+\biggl(\partial^{2}_{r}F\partial_{\theta}F\partial_{\theta}\Theta
			+\partial^{2}_{\theta}F\partial_{r}F\partial_{r}\Theta
			+2\partial_{r}\partial_{\theta}\Theta\partial_{\theta}F\partial_{r}F\biggr)
			\biggr\}=0
			\end{eqnarray}
where we defined dimensionless quantities according to $r:=\sqrt{\kappa_2/\kappa_4}\rho, \mu:=\kappa_0\kappa_4/\kappa_2^2$.

\begin{figure}[t]
\includegraphics[clip,height=30mm]{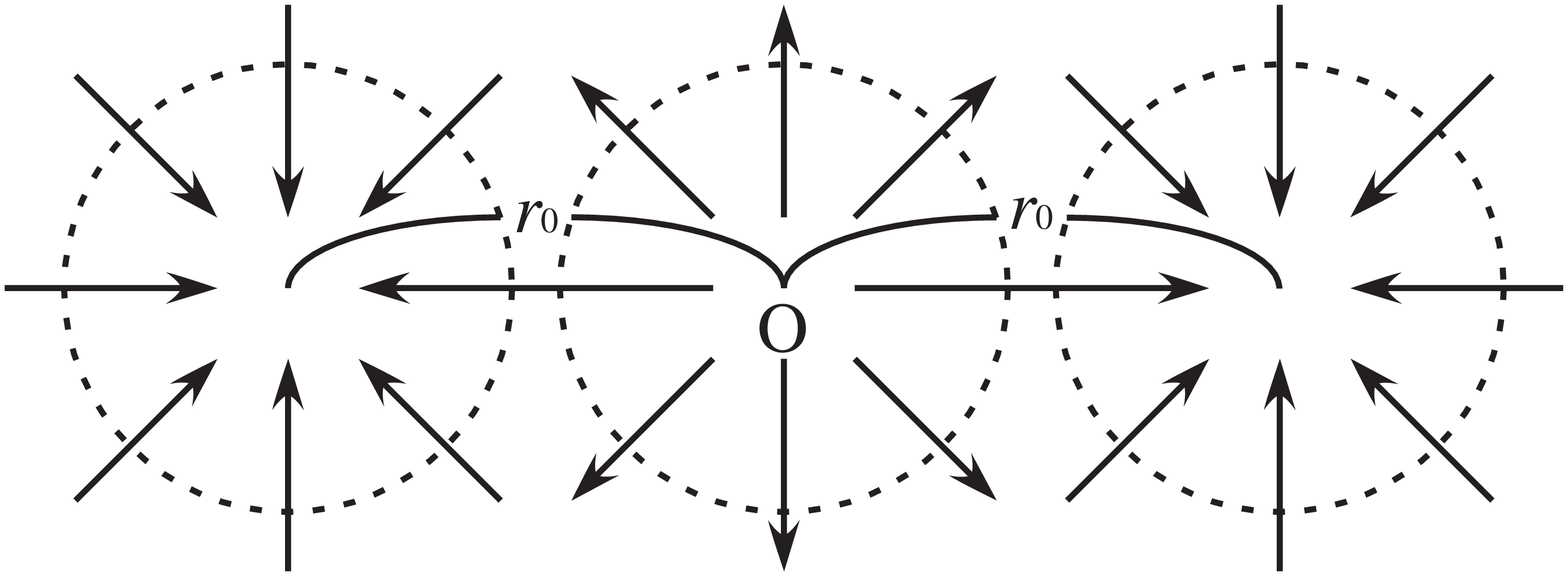}
\caption{\label{fig:cartoon}Cartoon figure of the arrow plot of $(n_1,n_2)$ in the case of three centered solution of the $B=3$. }
\end{figure}

\begin{figure*}[t]
\includegraphics[height=8.0cm, width=12.0cm]{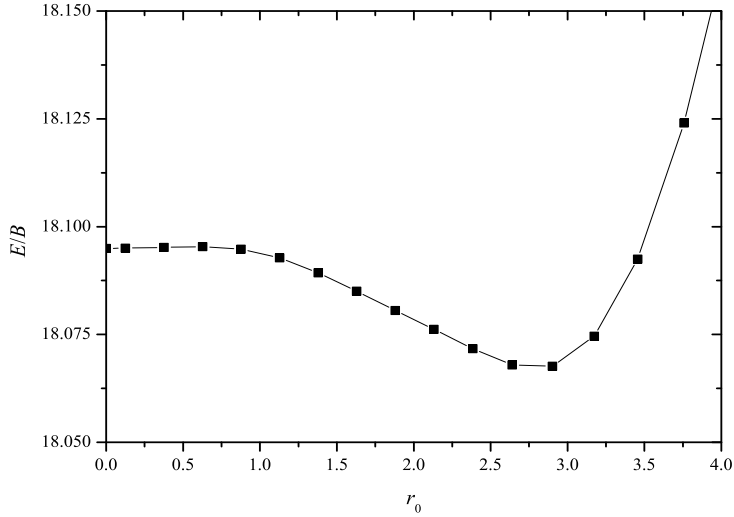}
\caption{\label{fig:energy_r0}The energy per topological charge of the $B=3$ as a function of the ``delocalization '' $r_0$, in the 
case of $\mu=0.08$. From the result, we extract $r_0= 2.6$ as the true distance of the centers. }
\end{figure*}

The static energy (\ref{senergy}) reduces to
\begin{eqnarray}
&&E_{{\rm static}}=E_2+E_4+E_0, \nonumber \\
&&E_2:=\int_{-\pi}^{\pi} d\theta
			\int_{0}^{\infty} rdr
			\biggl\{
			\frac{1}{2}\biggl((\partial_{r}F)^{2}
			+\sin^{2}F(\partial_{r}\Theta)^{2}\biggr)
			+\frac{1}{2r^{2}}\biggl(
			(\partial_{\theta}F)^{2}+\sin^{2}F(\partial_{\theta}\Theta)^{2}\biggr)\biggr\} \nonumber \\
&&E_4:=\int_{-\pi}^{\pi} d\theta
			\int_{0}^{\infty} rdr
    			\frac{1}{2r^{2}}\sin^{2}F\biggl(
			\partial_{r}F\partial_{\theta}\Theta-\partial_{\theta}F\partial_{r}\Theta
			\biggr)^2 \nonumber \\
&&E_0:=\int_{-\pi}^{\pi} d\theta
			\int_{0}^{\infty} rdr
			\mu V(\bm{n})
\end{eqnarray}
and the topological charge (\ref{topologicalcharge}) reduces to
\begin{eqnarray}
B=\int_{-\pi}^{\pi}d\theta\int_{0}^{\infty}rdr~b(r,\theta);~~~~
b(r,\theta):=-\frac{1}{4\pi r}\bigl[\partial_r (\cos F\partial_\theta \Theta)-\partial_\theta (\cos F\partial_r \Theta)\bigr].
\label{topologicalcharge_polar}
\end{eqnarray}

We introduce a 
deformation factor $\Delta$ in order to quantify the deformation of the solutions, defined as 
\begin{eqnarray}
\Delta^2 \equiv
\frac{2\pi}{B^2} \int_{-\pi}^{\pi}d\theta \left( \left[ \int_0^\infty rdr~
b(r,\theta)  \right]^2 \right)  -1.
\end{eqnarray}

\subsection{The $B=3$ solution}

The $\mathbb{Z}_2$-symmetry imposes that the equations are invariant under the transformations
\begin{eqnarray}
\theta \rightarrow \theta+\pi:~~
F(r,\theta)\rightarrow F(r,\theta+\pi),~~
\Theta_0 (r,\theta)\rightarrow \Theta_0(r,\theta+\pi)
\label{eq:symmetryft}
\end{eqnarray}
and at the axis of symmetry  $\theta=0$, 
\begin{eqnarray}
\theta \rightarrow -\theta:~~
F(r,\theta)\rightarrow F(r,-\theta),~~
\Theta_0 (r,\theta)\rightarrow -\Theta_0(r,-\theta).
\end{eqnarray}
Therefore we restrict our numerical calculations to the half-plane defined by
$0\leqq r \leqq \infty$ and $-\frac{\pi}{2}\leqq \theta\leqq \frac{\pi}{2}$. 

\begin{figure*}[t]
\includegraphics[height=6.0cm, width=9.0cm]{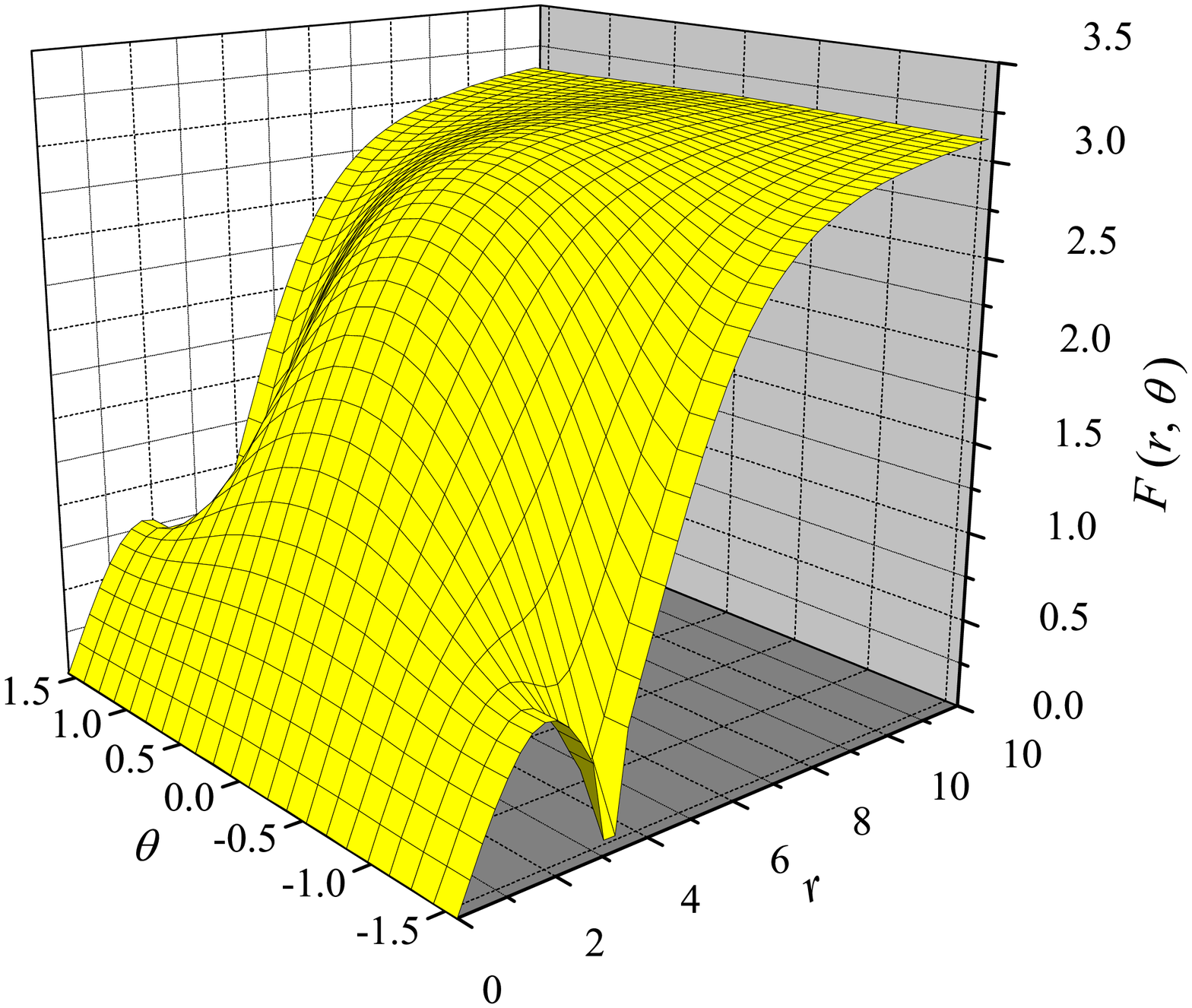}\hspace{-2cm}
\includegraphics[height=6.0cm, width=9.0cm]{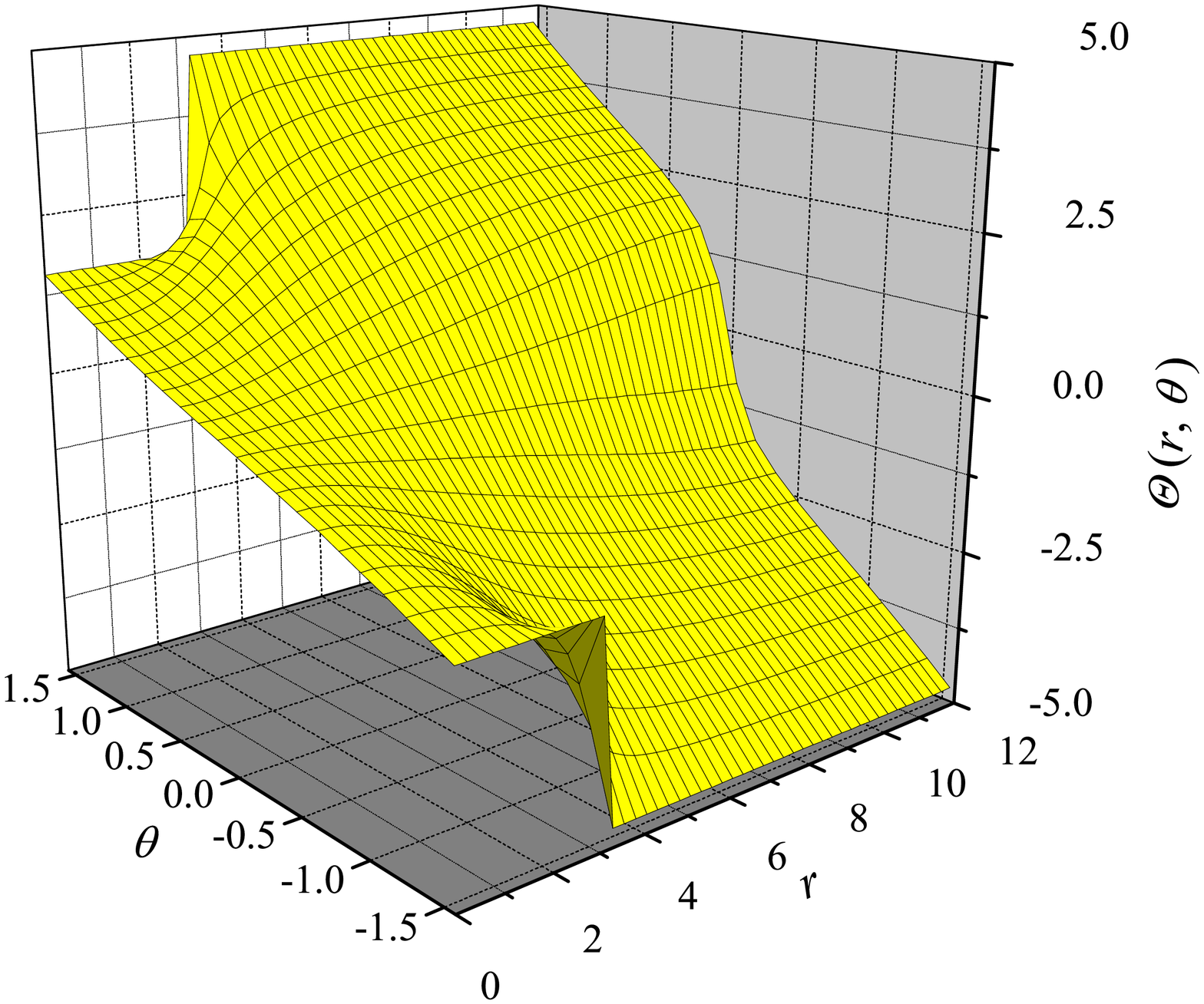}
\caption{\label{fig:profiles}The profiles $F(r,\theta)$, $\Theta(r,\theta)$ of the $B=3$ as solutions of 
the Euler-Lagrange equations with $\mu=0.08$. }
\end{figure*}

\begin{figure*}[t]
\includegraphics[width=4.5cm]{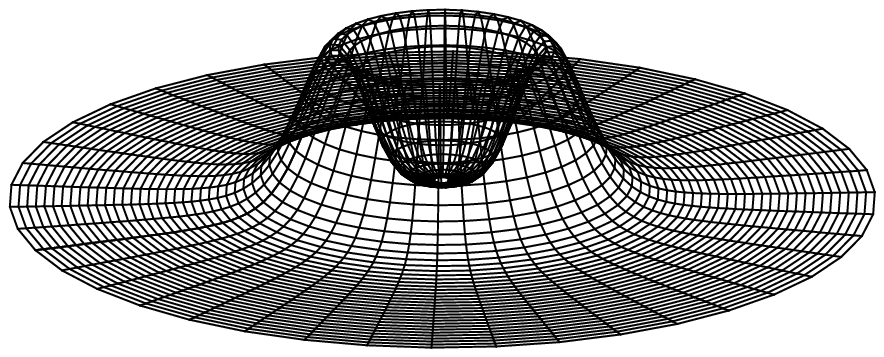}\hspace{-0.5cm}
\includegraphics[width=4.5cm]{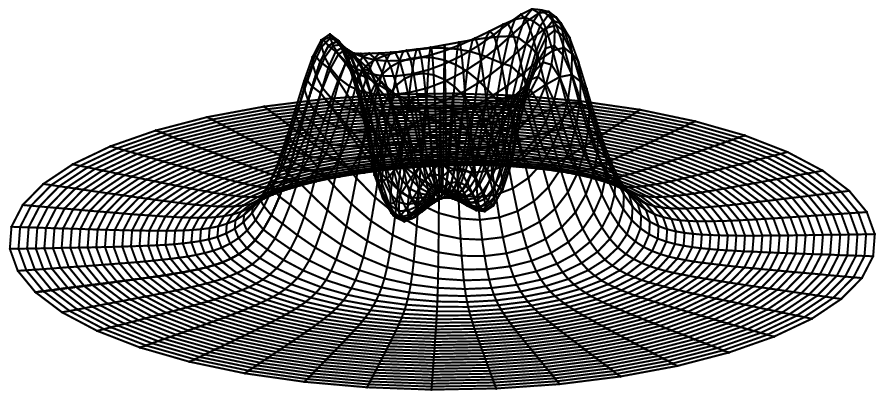}\hspace{-0.5cm}
\includegraphics[width=4.5cm]{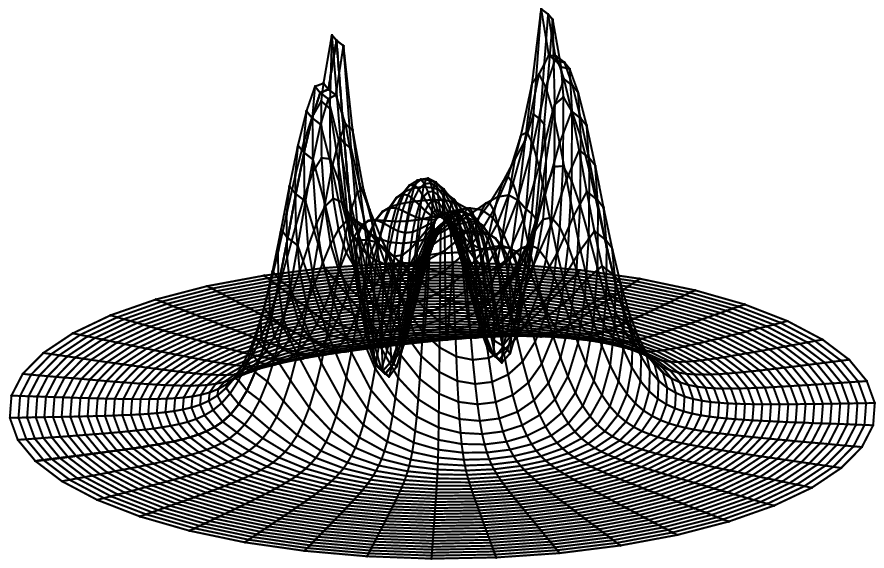}\hspace{-0.5cm}
\includegraphics[width=4.5cm]{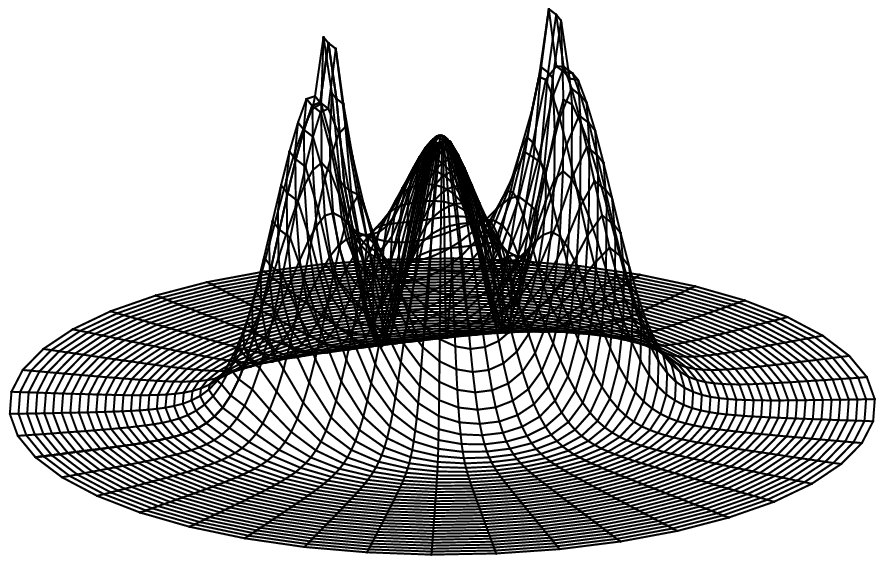}\\
{\footnotesize $\mu=0.04$\hspace{3cm}$\mu=0.06$\hspace{3cm}$\mu=0.12$\hspace{3cm}$\mu=0.64$} \\
\caption{\label{fig:energydensity1}The energy density of the $B=3$ solution.}
\end{figure*}

\begin{figure}[t]
\includegraphics[height=7.0cm, width=8.0cm]{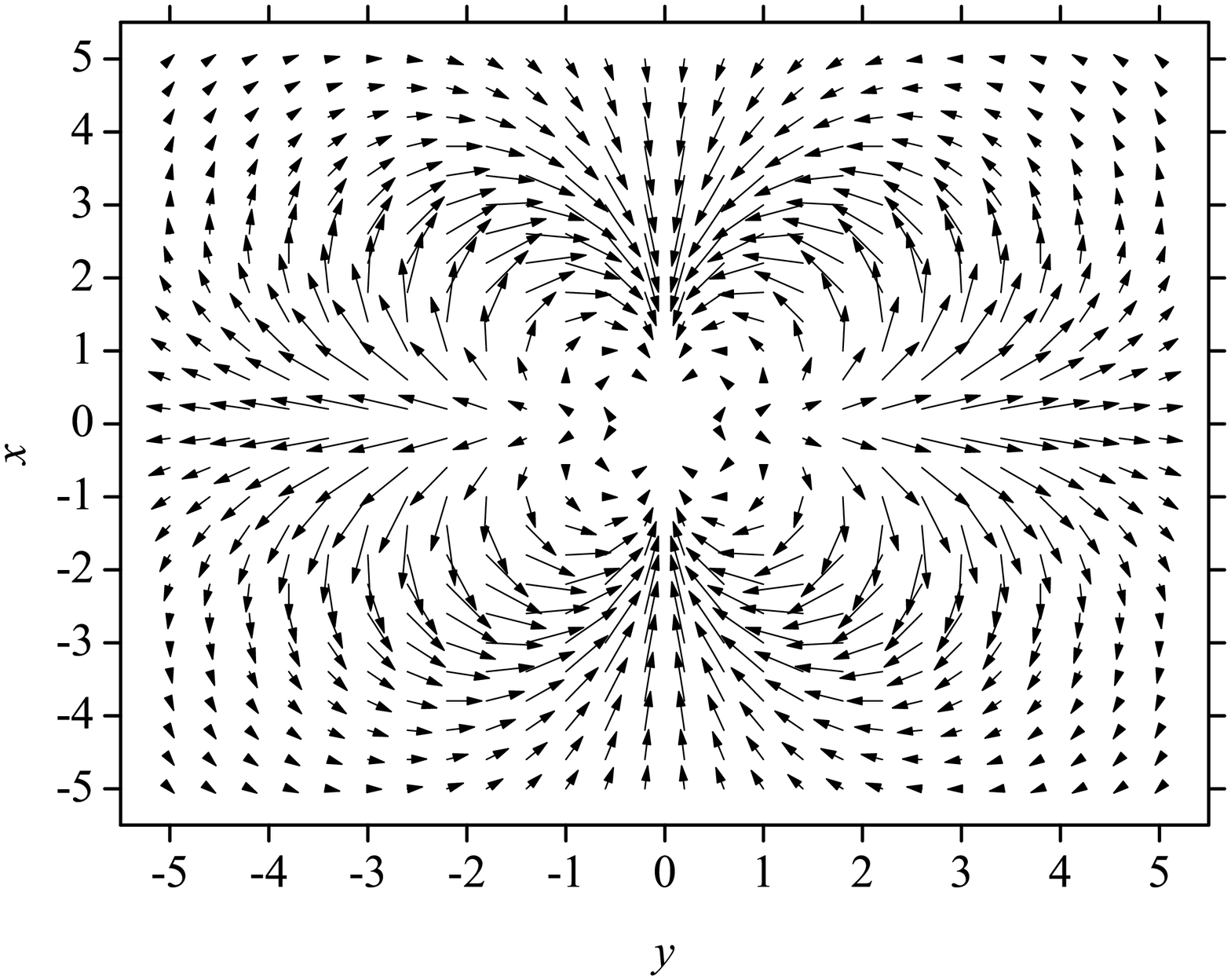}
\includegraphics[height=7.0cm, width=8.0cm]{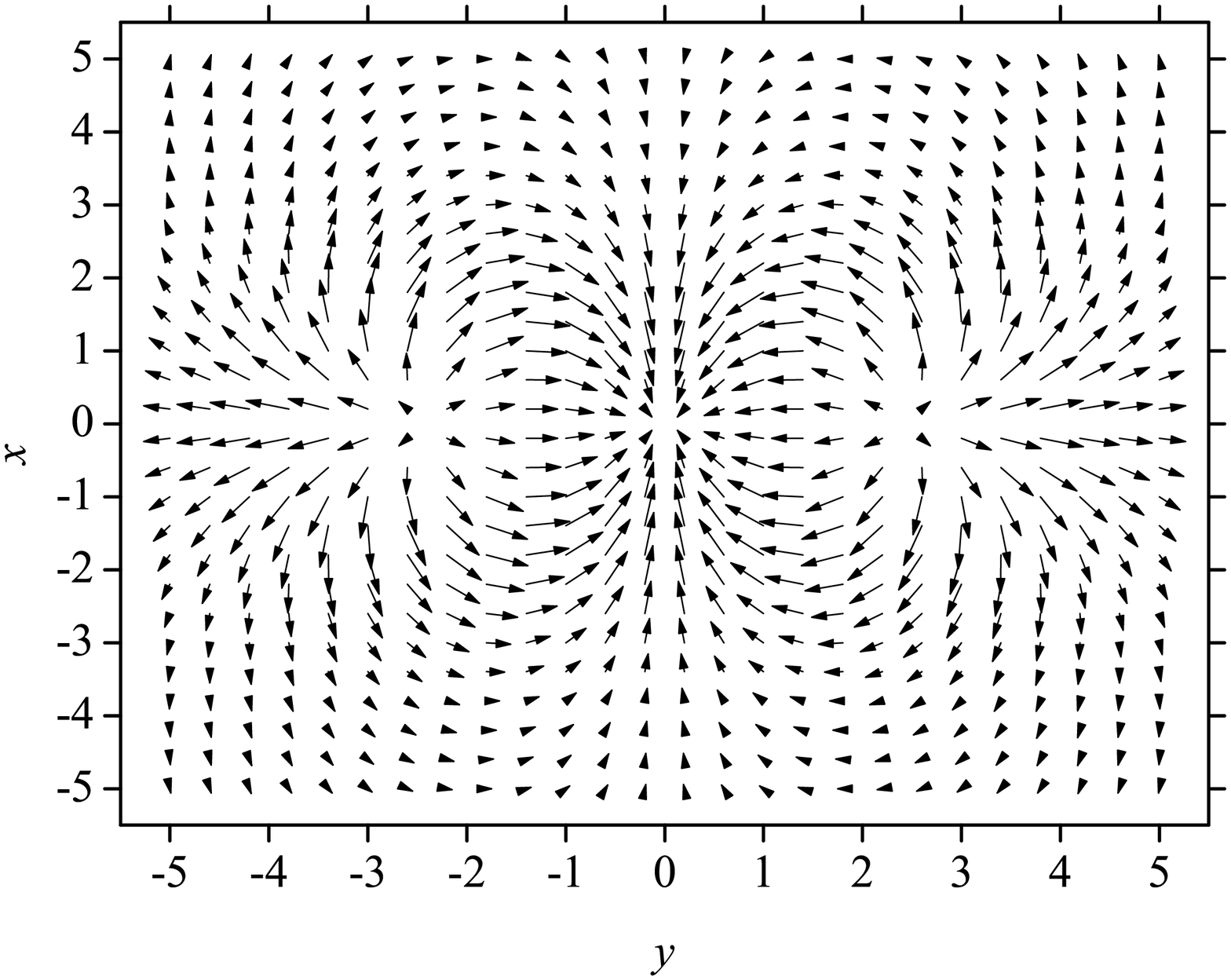}
\caption{\label{fig:arrowplot}The figures of the arrow plot of $\bm{n}$ of the $B=3$ solutions 
in the case of the one centered, $r_0=0$ solution (left) 
and the three centered, $r_0\sim 3.0$ solution (right) for $\mu=0.08$. 
The length of the arrow is relative to the norm of the vector, i.e. $\sqrt{n_1^2+n_2^2}$.}
\end{figure}

We tried using the following boundary conditions, 
\begin{eqnarray}
F(0,\theta)=0,~~
\partial_r\Theta(r,\theta)|_{r=0}=0,~~
F(\infty,\theta)=\pi,~~
\partial_r\Theta(r,\theta)|_{r=\infty}=0,~~
{\rm for}~~-\frac{\pi}{2}\leqq\theta\leqq\frac{\pi}{2}
\label{eq:boundarycondition00-r}
\end{eqnarray}
and 
\begin{eqnarray}
\partial_\theta F(r,\theta)|_{\theta=\pm\frac{\pi}{2}}=0,~~
\Theta(r,\pm\frac{\pi}{2})=\pm\frac{3}{2}\pi,~~
{\rm for}~~0\leqq r\leqq\infty.
\label{eq:boundarycondition00-t}
\end{eqnarray}
Although they are sufficient to get the solutions with axisymmetry (i.e.,the left panel of Fig.\ref{fig:annealing_e}), they are not suited for the symmetry breaking solutions and additional conditions seem to be required to reproduce the results.
The reason is that for a critical value of $\mu$, peaks appear and grow at a finite distance from the origin~(see Fig.\ref{fig:annealing_e}). 
These points are singular points where the direction of the component arrow of $\bm{n}$ is frustrated. 
The geometric structure is schematically drawn in Fig.\ref{fig:cartoon}; 
one easily see that for $r<r_0$ (here $r$ is the distance from the origin), 
the geometry looks like the one center hedgehog, while for $r>r_0$ the arrow winds three times around the origin. 
We take this property of the solution geometry into account by using the following boundary conditions (\ref{eq:boundarycondition00-r}), 
\begin{eqnarray}
&&\partial_\theta F(r,\theta)|_{\theta=\pm\frac{\pi}{2}}=0,~~
{\rm for}~~0\leqq r < r_{0},\ r_{0} < r\leqq\infty, 
\label{eq:boundarycondition01-r1}\\
&&F(r_0,\pm\frac{\pi}{2})=0
\label{eq:boundarycondition01-r2}
\end{eqnarray}
and
\begin{eqnarray}
\Theta(r,\pm\frac{\pi}{2})=\left\{
\begin{array}{lrr}
\displaystyle \pm\frac{\pi}{2}\ & 0\leqq r < & r_{0}  
\\\\
\displaystyle \pm\frac{3}{2}\pi\ & r_{0} \leqq r\leqq & \infty 
\end{array}
\right.
\label{eq:boundarycondition01-t}
\end{eqnarray}
where $r_0$ is a constant describing the center distance of the solution constituent solitons.
Note that when $r_0\to 0$ the boundary conditions (\ref{eq:boundarycondition01-r1})-(\ref{eq:boundarycondition01-t})
coincide with the case of the axisymmetry (\ref{eq:boundarycondition00-t}).
One can easily check that these modified boundary conditions correctly reproduce $B=3$~(see appendix).

The equations with the appropriate boundary conditions can be solved numerically by the SOR method. 
In order to get the non-axisymmetric solution, we employ the following procedures.
We begin the SOR for a fixed $r_0$ with following initial conditions which satisfies the boundary behavior of the axisymmetric solutions
\begin{eqnarray}
F(r,\theta)|_{\rm initial}=\pi (1-e^{-Ar^3}),~~~~
\Theta(r,\theta)|_{\rm initial}=3\theta.
\end{eqnarray}
For changing $r_0$, we repeat the computation and finally reach the true value of $r_0$ by minimizing the energy par 
topological charge $E/B$, which realizes the true boundary condition for the system. 
In terms of the procedure, we successfully find the non-axisymmetric shape of the solutions.

In Fig.\ref{fig:energy_r0} we plot a typical $E/B$ as a function of $r_0$. In this case, we could choose $r_0=2.6$ as a minimized energy per unit charge. 
Fig.\ref{fig:profiles} shows a numerical result of the profile functions $F(\rho,\theta),\Theta(\rho,\theta)$.
In Fig.\ref{fig:energydensity1}, we present the corresponding topological charge density for several potential parameter $\mu$. 
One easily sees that for increasing values of $\mu$, the peaks grow and a new peak appears at the origin. 
Note that the $F,\Theta$ profiles contain singularities/discontinuities corresponding to the centers of the constituents. In Fig.\ref{fig:arrowplot} we plot the configurations of $\bm{n}$, i.e., the vectors $(n_1,n_2)$ on the 
plane for the case of $r_0=0.0$ and $r_0\sim 3.0$ which corresponds to minima of the case of $\mu=0.08$. 
One easily see that these are perfectly regular. 

\begin{figure*}
\hspace{-1cm}
\includegraphics[height=8.0cm, width=12.0cm]{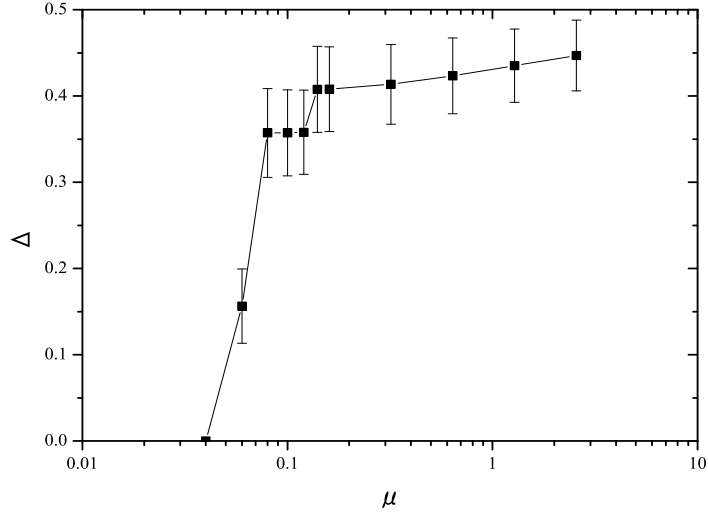}
\caption{\label{fig:deform}The deformation parameter $\Delta$ of the $B=3$ solution as a function of $\mu$.}
\end{figure*}

In Fig.\ref{fig:deform} we plot the behavior of the deformation parameter $\Delta$ for changing the strength of the potential. 
For larger $\mu$, the deformation monotonically grows. Note that there is a critical point of the deformation of the solution: for $\mu\sim 0.04$, deformed solutions suddenly appear. The  error-bar is due to the fact that the results involve a systematic uncertainly for the determination of $r_0$.

\begin{figure*}
\hspace{-1cm}
\includegraphics[height=6.8cm, width=9.3cm]{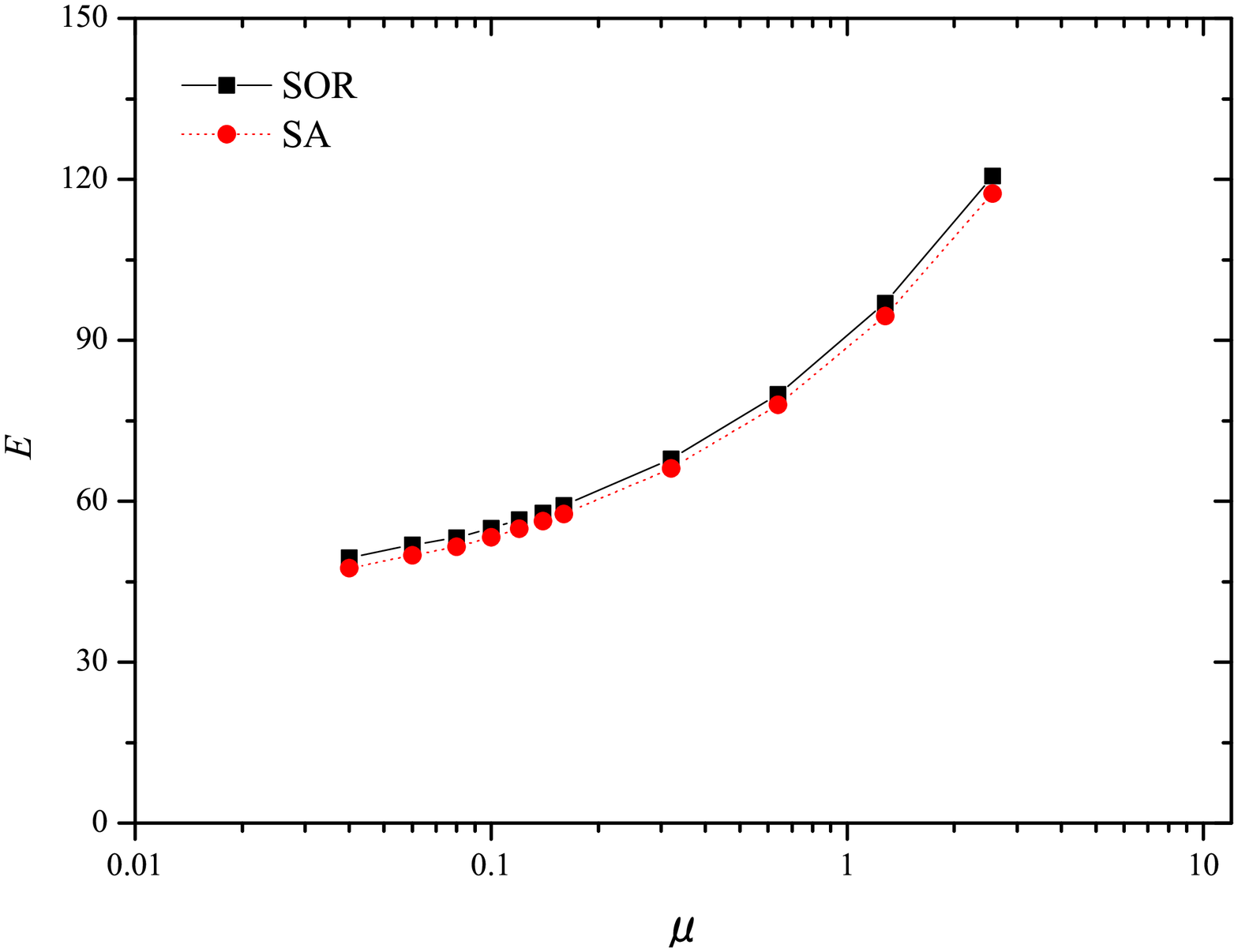}
\includegraphics[height=6.8cm, width=9.3cm]{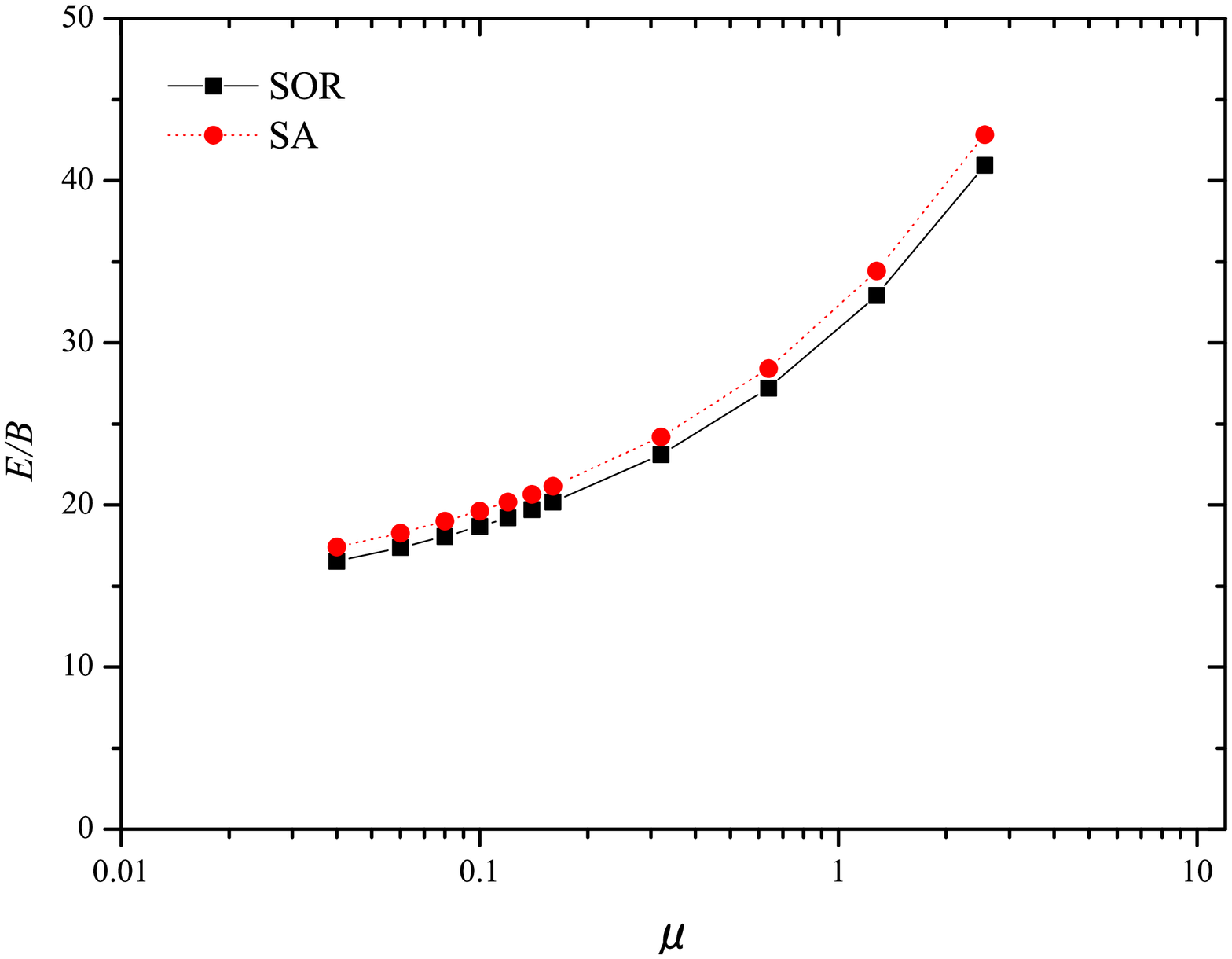}
\caption{\label{fig:mu-e}The comparison of the static energies (left) and the energies per the topological charge (right) 
of the $B=3$ solutions for the successive over-relaxation (straight line) and the simulated annealing (dotted line).}
\end{figure*}

The solution of the Euler-Lagrange equation should coincide with result of the energy-minimization scheme. 
In Fig.\ref{fig:mu-e}, we compare the energy $E$ and also the energy per topological charge $E/B$ in terms of the SOR and of the SA for the potential parameters $\mu$. 
Both methods contain uncertainties in several reasons, in particular the SA always tends to underestimate 
the topological charge as well as the energy. As a result, the energy of the SA is lower value than that of the SOR while the energy per topological charge is slightly higher, but the results are in good agreement within $\sim 5\%$.

For higher charge, solutions breaking the axisymmetry appear even for very small $\mu$. 

\subsection{The $B=4$ solution}
The cases with a $\mathbb{Z}_2$-symmetry are treated essentially in the same way as in the $B=3$ case.
Therefore, here we concentrate on the case of $\mathbb{Z}_3$-symmetry. A typical result of the energy density is shown in Fig. \ref{fig:annealing_e4}.

For the  $\mathbb{Z}_3$-symmetric case, the equations must be invariant under the transformations
\begin{eqnarray}
\theta \rightarrow \theta+\frac{2\pi}{3}:~~
F(r,\theta)\rightarrow F(r,\theta+\frac{2\pi}{3}),~~
\Theta_0 (r,\theta)\rightarrow \Theta_0(r,\theta+\frac{2\pi}{3})
\end{eqnarray}
and if we assume that one of the peak is located on the $x$-axis ($\theta=\pi$), 
the equations are invariant under the transformations
\begin{eqnarray}
\theta \rightarrow -\theta:~~
F(r,\theta)\rightarrow F(r,-\theta),~~
\Theta_0 (r,\theta)\rightarrow -\Theta_0(r,-\theta).
\end{eqnarray}
Therefore we perform our numerical calculations on the one third-plane defined as
$0\leqq r \leqq \infty$ and $-\frac{\pi}{3}\leqq \theta\leqq \frac{\pi}{3}$. The functions on the 
whole plane can be estimated in terms of the above symmetry. 

We use the following boundary conditions:
\begin{eqnarray}
F(0,\theta)=0,~~
\partial_r\Theta(r,\theta)|_{r=0}=0,~~
F(\infty,\theta)=\pi,~~
\partial_r\Theta(r,\theta)|_{r=\infty}=0,~~
{\rm for}~~-\frac{\pi}{3}\leqq\theta\leqq\frac{\pi}{3}
\end{eqnarray}
and 
\begin{eqnarray}
\partial_\theta F(r,\theta)|_{\theta=\pm\frac{2\pi}{3}}=0,~~
\Theta(r,\pm\frac{\pi}{3})=\pm\frac{4\pi}{3},~~
{\rm for}~~0\leqq r\leqq\infty
\end{eqnarray}
for the axial symmetric solutions and when the constituent soliton centers are located on the 
line of $\theta=\pm \frac{\pi}{3}, \pi$ , we take the singularities of the centers into account by imposing
\begin{eqnarray}
F(0,\theta)=0,~~
\partial_r\Theta(r,\theta)|_{r=0}=0,~~
F(\infty,\theta)=\pi,~~
\partial_r\Theta(r,\theta)|_{r=\infty}=0,~~
{\rm for}~~-\frac{\pi}{3}\leqq\theta\leqq\frac{\pi}{3}
\end{eqnarray}
\begin{eqnarray}
&&\partial_\theta F(r,\theta)|_{\theta=\pm\frac{\pi}{3}}=0,~~
{\rm for}~~0\leqq r < r_{0},\ r_{0} < r\leqq\infty, 
\\
&&F(r_0,\pm\frac{\pi}{3})=0
\end{eqnarray}
and 
\begin{eqnarray}
\Theta(r,\pm\frac{\pi}{3})=\left\{
\begin{array}{lrr}
\displaystyle \pm\frac{\pi}{3}\ & 0\leqq r < & r_{0}  
\\\\
\displaystyle \pm\frac{4\pi}{3}\ & r_{0} \leqq r\leqq & \infty.
\end{array}
\right.
\end{eqnarray}
Again for $r_0\to 0$ the boundary conditions reduces to the axisymmetric case.

\begin{figure*}[t]
\includegraphics[height=6.0cm, width=9.0cm]{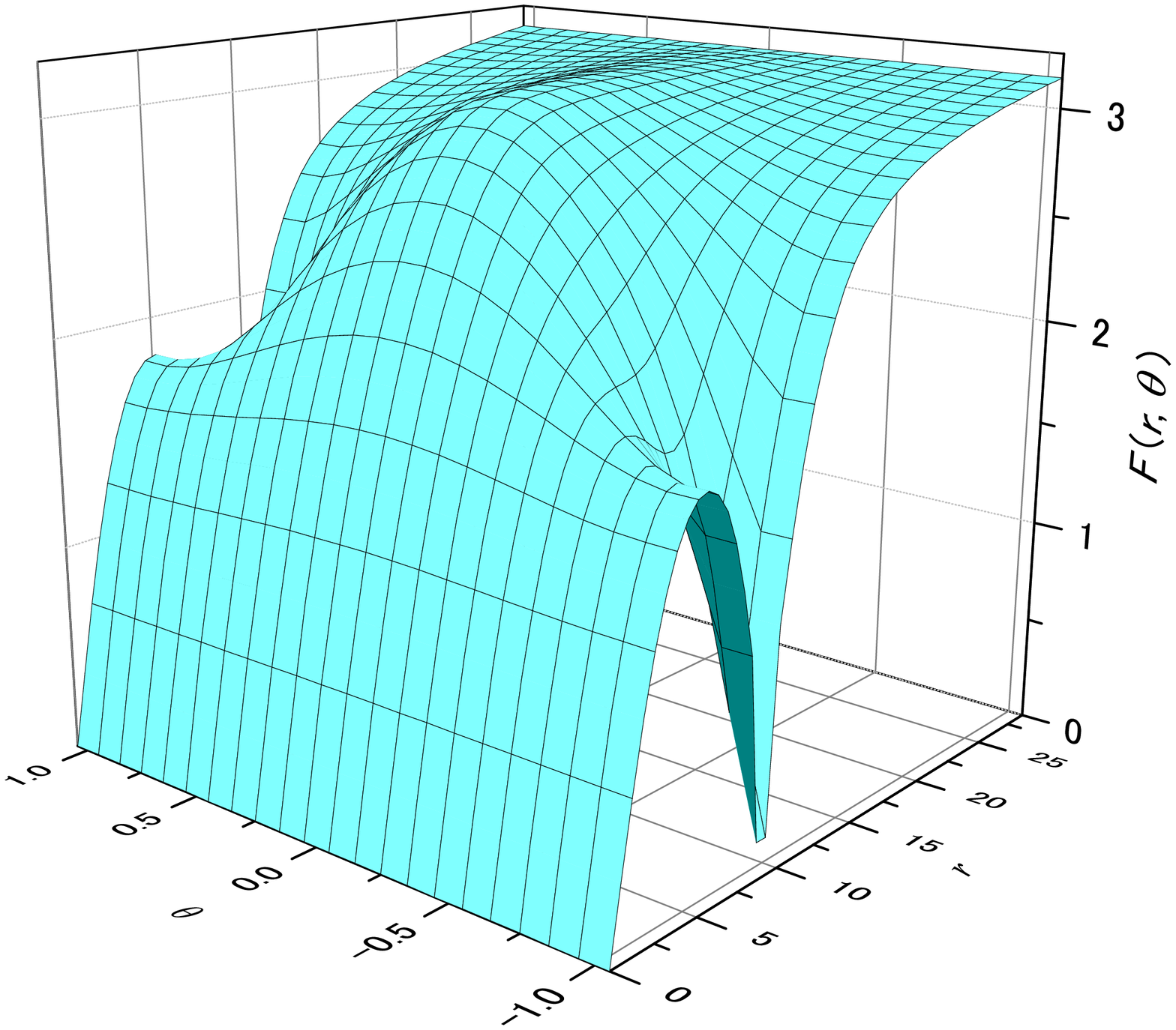}\hspace{-2cm}
\includegraphics[height=6.0cm, width=9.0cm]{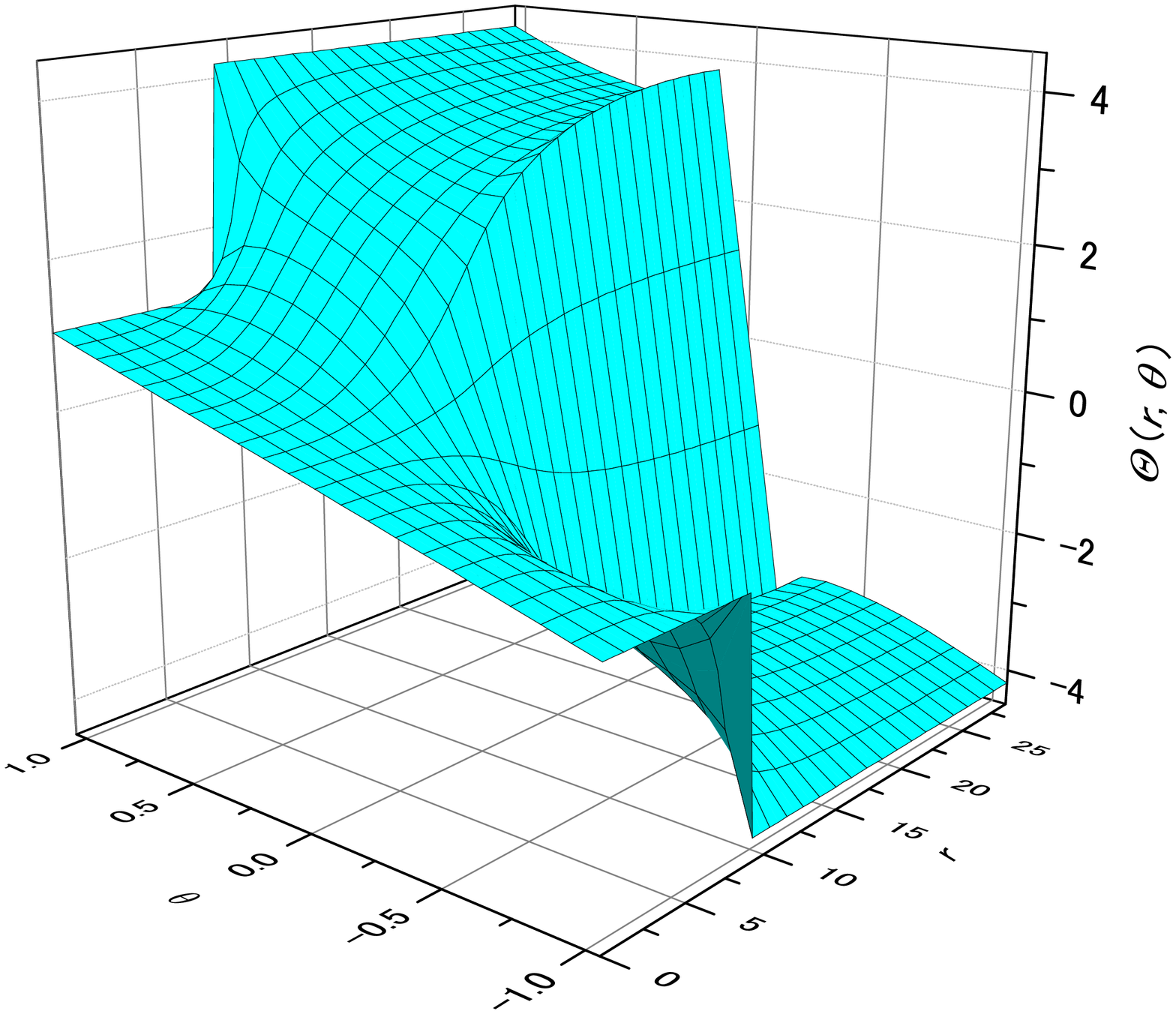}
\caption{\label{fig:profiles4}The profiles $F(r,\theta)$, $\Theta(r,\theta)$ of the $B=4$ as solutions of 
the Euler-Lagrange equations with $s=2,\mu=0.01$. }
\end{figure*}

Fig.\ref{fig:profiles4} shows a numerical result of the profile functions $F(\rho,\theta),\Theta(\rho,\theta)$.

\section{The gravitating baby-Skyrme model in six dimensions}
We are now are ready to include gravity to the baby-Skyrme brane model. In this paper, we will focus on the $B=3$ but
the method is straightforwardly extendable to the cases of different $B$. We also show the result for $B=4$ as an example.
  
The total action for the gravitating baby-Skyrme model in six dimensions is of the form $S = S_{\text{grav}} + S_{\text{baby}}$ with
\begin{equation}
S_{\text{grav}} = \int d^{6}x\sqrt{-g}\left( \frac{1}{2\chi_{(6)}}R - \Lambda_{(6)} \right), 
\label{eq:gravity-action}
\end{equation}
\begin{eqnarray}
S_{\text{baby}} = \int d^{6}x\sqrt{-g}{\cal L}_{\text{baby}},
\label{eq:brane-action}
\end{eqnarray}
where $\Lambda_{(6)}$ is the bulk cosmological constant, and $\chi_{(6)} = 8\pi G_{(6)} = 8\pi/M_{(6)}^{4}$.

The coefficients $\kappa_{2,4,0}$ in Eq.(\ref{eq:brane-action}) are the dimensionful coupling constants in the gravitating baby-Skyrme model with dimensions $[\Lambda_{(6)}] = M^{6}$, $[\chi_{(6)}] = M^{-4}$,
$[\kappa_{2}] = M^{4}$, $[\kappa_{4}] = M^{2}$ $[\kappa_{0}] = M^{6}$.

\subsection{The Ansatz}
We consider the case of axial symmetry breaking in the extra dimensions, 
and we use the following metric ansatz inspired by the Lewis-Papapetrou ansatz in $4$-dimensions
\begin{equation}
	ds^{2} = M^{2}(\rho,\theta)g_{\mu\nu}^{(4)}dx^{\mu}dx^{\nu} - \frac{L^2(\rho,\theta)}{M^2(\rho,\theta)}(d\rho^{2} +\rho^2d\theta^{2})
	\label{eq:6D-metric}
\end{equation}
where $\rho \in [0, \infty)$ and $\theta \in [ -\pi,\pi]$ are the coordinates associated with the extra dimensions.
We further model a cosmological constant on the brane by considering the following form of the four dimensional 
subspace (described by $g_{\mu\nu}^{(4)}$ in Eq.\eqref{eq:6D-metric})
\begin{equation}
	ds_{(4)}^{2} = g_{\mu\nu}^{(4)}dx^{\mu}dx^{\nu} = dt^{2} - \delta_{ij}e^{2H(t)}dx^{i}dx^{j}.
	\label{eq:4D-metric}
\end{equation}
$H(t)$ is a function of the time coordinate $t$ which describes the inflation in the four dimensions.

\subsection{Field equations of the model}
In order to rewrite the system in terms of dimensionless quantities, we define the dimensionless variable
\begin{equation}
	r := \sqrt{\frac{\kappa_{2}}{\kappa_{4}}}\,\rho,\;\;\;
	\mathcal{H}(t) := \sqrt{\frac{\kappa_{4}}{\kappa_{2}}}H(t).
	\label{eq:dimensionless}
\end{equation}

and dimensionless parameters
\begin{eqnarray}
\alpha := \chi_{(6)}\kappa_{2},\ \beta := \Lambda_{(6)}\frac{\kappa_{4}}{\kappa_{2}^{2}},\ \mu :=\kappa_{0}\frac{\kappa_{4}}{\kappa_{2}^{2}}.
\end{eqnarray}

After straightforward computations, the Einstein equations reduce to
\begin{eqnarray}
&&\frac{M^2}{L^2}\Bigl[
2\frac{\partial_r^2 M}{M}+4\Bigl(\frac{\partial_r M}{M}\Bigr)^2+\frac{2}{r}\frac{\partial_rM}{M}
+ \frac{\partial_r^2 L}{L}-\Bigl(\frac{\partial_r L}{L}\Bigr)^2+\frac{1}{r}\frac{\partial_rL}{L}
\Bigr] \nonumber \\
&&\hspace{1cm}+\frac{M^2}{r^2L^2}\Bigl[
2\frac{\partial_\theta^2 M}{M}
+4\Bigl(\frac{\partial_\theta M}{M}\Bigr)^2
+ \frac{\partial_\theta^2 L}{L}
-\Bigl(\frac{\partial_\theta L}{L}\Bigr)^2
\Bigr]
- \frac{3\mathcal{H}_{,t}^{2}}{M^{2}} = \alpha( \tau_{0} - \beta ),
		\label{eq:Einstein-eq_t} \\
&&\frac{M^2}{L^2}\Bigl[
2\frac{\partial_r^2 M}{M}+4\Bigl(\frac{\partial_r M}{M}\Bigr)^2+\frac{2}{r}\frac{\partial_rM}{M}
+ \frac{\partial_r^2 L}{L}-\Bigl(\frac{\partial_r L}{L}\Bigr)^2+\frac{1}{r}\frac{\partial_rL}{L}
\Bigr] \nonumber \\
&&\hspace{1cm}+\frac{M^2}{r^2L^2}\Bigl[
2\frac{\partial_\theta^2 M}{M}
+4\Bigl(\frac{\partial_\theta M}{M}\Bigr)^2
+ \frac{\partial_\theta^2 L}{L}
-\Bigl(\frac{\partial_\theta L}{L}\Bigr)^2
\Bigr]
-\frac{2\mathcal{H}_{,t,t}+3\mathcal{H}_{,t}^{2}}{M^{2}} = \alpha( \tau_{0} - \beta ),
\label{eq:Einstein-eq_3} \\
&&\frac{M^2}{L^2}\Bigl[
2\Bigl(\frac{\partial_r M}{M}\Bigr)^2 +\frac{4}{r}\frac{\partial_rM}{M}+4\frac{\partial_r M\partial_r L}{ML}
\Bigr] \nonumber \\
&&\hspace{1cm}+\frac{M^2}{r^2L^2}\Bigl[4\frac{\partial_\theta^2 M}{M}+10\Bigl(\frac{\partial_\theta M}{M}\Bigr)^2
-4\frac{\partial_\theta M\partial_\theta L}{ML}\Bigr] 
- \frac{3\mathcal{H}_{,t,t}+6\mathcal{H}_{,t}^{2}}{M^{2}}= \alpha( \tau_{r} - \beta ),
\label{eq:Einstein-eq_r} \\
&&\frac{M^2}{L^2}\Bigl[
4\frac{\partial_r^2 M}{M} + 10\Bigl(\frac{\partial_r M}{M}\Bigr)^2 
-4\frac{\partial_r M\partial_r L}{ML}\Bigr] \nonumber \\
&&\hspace{1cm}+\frac{M^2}{r^2L^2}\Bigl[
2\Bigl(\frac{\partial_\theta M}{M}\Bigr)^2+4\frac{\partial_\theta M\partial_\theta L}{ML}\Bigr]
- \frac{3\mathcal{H}_{,t,t}+6\mathcal{H}_{,t}^{2}}{M^{2}}= \alpha( \tau_{\theta} - \beta ),
\label{eq:Einstein-eq_th}
\end{eqnarray}
where we used the notation $\mathcal{H}_{,t} := \partial_{t}\mathcal{H}(t)$ and $\mathcal{H}_{,t,t} := \partial_{t}^{2}\mathcal{H}(t)$.
Let us note that (\ref{eq:Einstein-eq_t}) and (\ref{eq:Einstein-eq_3}) are the four-dimensional components of the Einstein equation, 
while (\ref{eq:Einstein-eq_r}) and (\ref{eq:Einstein-eq_th}) are the extra-dimensional components.
The components of the dimensionless energy-momentum (EM) tensor in Eqs.(\ref{eq:Einstein-eq_t})-(\ref{eq:Einstein-eq_th}) are given by
\begin{equation}
\begin{aligned}
	&\tau_{0} = - \frac{M^2}{2L^2}\bigl((\partial_r F)^2+\sin^2F(\partial_r\Theta)^2\bigr)
	-\frac{M^2}{2r^2L^2}\bigl((\partial_\theta F)^2+\sin^2F(\partial_\theta\Theta)^2\bigr) 
	-\frac{M^4}{2r^2L^4}\sin ^2 F(\partial_{[r} F\partial_{\theta]}\Theta)^2- \mu V(\bm{n}),\\
	&\tau_{r} =~~~\frac{M^2}{2L^2}\bigl((\partial_r F)^2+\sin^2F(\partial_r\Theta)^2\bigr)
	-\frac{M^2}{2r^2L^2}\bigl((\partial_\theta F)^2+\sin^2F(\partial_\theta\Theta)^2\bigr) 
	+\frac{M^4}{2r^2L^4}\sin ^2 F(\partial_{[r} F\partial_{\theta]}\Theta)^2- \mu V(\bm{n}),\\
	&\tau_{\theta} =- \frac{M^2}{2L^2}\bigl((\partial_r F)^2+\sin^2F(\partial_r\Theta)^2\bigr)
	+\frac{M^2}{2r^2L^2}\bigl((\partial_\theta F)^2+\sin^2F(\partial_\theta\Theta)^2\bigr) 
	+\frac{M^4}{2r^2L^4}\sin ^2 F(\partial_{[r} F\partial_{\theta]}\Theta)^2- \mu V(\bm{n}),
\label{emtensor}
\end{aligned}
\end{equation}
where $\partial_{[r} F\partial_{\theta]}\Theta:=\partial_rF\partial_\theta\Theta-\partial_\theta F\partial_r\Theta$.

From (\ref{eq:Einstein-eq_t}),(\ref{eq:Einstein-eq_3}) one directly sees that the function $\mathcal{H}(t)$ must be linear in time:
\begin{eqnarray}
\mathcal{H}(t)=\mathcal{H}_0t
\end{eqnarray}
where $\mathcal{H}_0$ is a constant called Hubble parameter. For later convenience, we introduce a dimensionless parameter
\begin{eqnarray}
\gamma:=\mathcal{H}_0^2=\frac{\kappa_4}{\kappa_2}H_0^2.
\end{eqnarray}
We finally get the following equations by using a suitable linear combination of  (\ref{eq:Einstein-eq_t})-(\ref{eq:Einstein-eq_th})
\begin{eqnarray}
&&\frac{M^2}{L^2}\Bigl[2\frac{\partial_r^2M}{M}+6\Bigl(\frac{\partial_rM}{M}\Bigr)^2+\frac{2}{r}\frac{\partial_rM}{M}\Bigr]
+\frac{M^2}{r^2L^2}\Bigl[2\frac{\partial_\theta^2M}{M}+6\Bigl(\frac{\partial_\theta M}{M}\Bigr)^2\Bigr]
-\frac{6\gamma}{M^2}=\alpha (\frac{1}{2}\tau_r+\frac{1}{2}\tau_\theta-\beta) 
\label{eq:Einstein-eq_m}\\
&&\frac{M^2}{L^2}\Bigl[\frac{\partial_r^2L}{L}-\Bigl(\frac{\partial_rL}{L}\Bigr)^2+\frac{1}{r}\frac{\partial_rL}{L}-2\Bigl(\frac{\partial_r M}{M}\Bigr)^2\Bigr]
\nonumber \\
&&\hspace{4cm}+\frac{M^2}{r^2L^2}\Bigl[\frac{\partial_\theta^2L}{L}-\Bigl(\frac{\partial_\theta L}{L}\Bigr)^2-2\Bigl(\frac{\partial_\theta M}{M}\Bigr)^2\Bigr]
+\frac{3\gamma}{M^2}=\alpha (\tau_0-\frac{1}{2}\tau_r-\frac{1}{2}\tau_\theta). 
\label{eq:Einstein-eq_L}
\end{eqnarray}

For the matter fields, the equations are given by
\begin{eqnarray}
&&\frac{M^2}{L^2}\Bigl\{\partial_r^2 F
+\Bigl(4\frac{\partial_r M}{M}+\frac{1}{r}\Bigr)\partial_r F
-\frac{1}{2}\sin 2F (\partial_r\Theta)^2\Bigr\}
+\frac{M^2}{r^2L^2}\Bigl(\partial_\theta^2 F
+4\frac{\partial_\theta M}{M}
\partial_\theta F
-\frac{1}{2}\sin 2F (\partial_\theta\Theta)^2
\Bigr) \nonumber \\
&&+\frac{M^4}{r^2L^4}\sin^2 F\Bigl[
\Bigl\{
\Bigl(6\frac{\partial_r M}{M}-2\frac{\partial_r L}{L}-\frac{1}{r}\Bigr)\partial_\theta \Theta
-\Bigl(6\frac{\partial_\theta M}{M}-2\frac{\partial_\theta L}{L}\Bigr)\partial_r \Theta\Bigr\}
(\partial_{[r}F\partial_{\theta ]}\Theta)
\nonumber \\
&&\hspace{1.0cm}+
\partial_r^2 F(\partial_\theta \Theta)^2+\partial_\theta^2F(\partial_r\Theta)^2
+(\partial_{\{r} F\partial_{\theta\}}\Theta)\partial_{r\theta}\Theta \nonumber \\
&&\hspace{1.0cm}-(\partial_r^2\Theta\partial_\theta\Theta\partial_\theta F
+\partial_\theta^2\Theta\partial_r\Theta\partial_rF+2\partial_{r\theta}F \partial_\theta\Theta \partial_r\Theta)
+\cot F (\partial_{\{r}F\partial_{\theta\}}\Theta)^2
\Bigr]-\mu\frac{\partial V}{\partial F}=0 
\label{eq:equationf}  \\ \nonumber \\
&&\frac{M^2}{L^2}\Bigl\{
\partial_r^2 \Theta+\Bigl(4\frac{\partial_r M}{M}+\frac{1}{r}\Bigr)\partial_r \Theta
+2\cot F\partial_rF\partial_r \Theta\Bigl\}
+\frac{M^2}{r^2L^2}\Bigl(\partial_\theta^2 \Theta+4\frac{\partial_\theta M}{M}
\partial_\theta \Theta
+2\cot F\partial_\theta F\partial_\theta \Theta\Bigr) 
\nonumber\\
&&-\frac{M^4}{r^2L^4}\Bigl[
\Bigl\{
\Bigl(6\frac{\partial_r M}{M}-2\frac{\partial_r L}{L}-\frac{1}{r}\Bigr)\partial_\theta F
-\Bigl(6\frac{\partial_\theta M}{M}-2\frac{\partial_\theta L}{L}\Bigr)\partial_r F\Bigr\}
(\partial_{[r}F\partial_{\theta ]}\Theta) 
\nonumber \\
&&\hspace{1.0cm}
+\partial_r^2 \Theta(\partial_\theta F)^2+\partial_\theta^2\Theta(\partial_rF)^2
+(\partial_{\{r }F\partial_{\theta\} }\Theta )\partial_{r\theta}F \nonumber \\
&&\hspace{1.5cm}-(\partial_r^2F\partial_\theta F\partial_\theta \Theta
+\partial_\theta^2F\partial_rF\partial_r\Theta+2\partial_{r\theta}\Theta \partial_\theta F \partial_rF)
\Bigr]=0
\label{eq:equationt}
\end{eqnarray}
where $\partial_{\{r }F\partial_{\theta\} }\Theta:=\partial_rF\partial_\theta\Theta+\partial_\theta F\partial_r\Theta$ and $\partial_{r\theta}:=\partial_r\partial_\theta $.

\begin{figure*}
\includegraphics[clip,width=80mm]{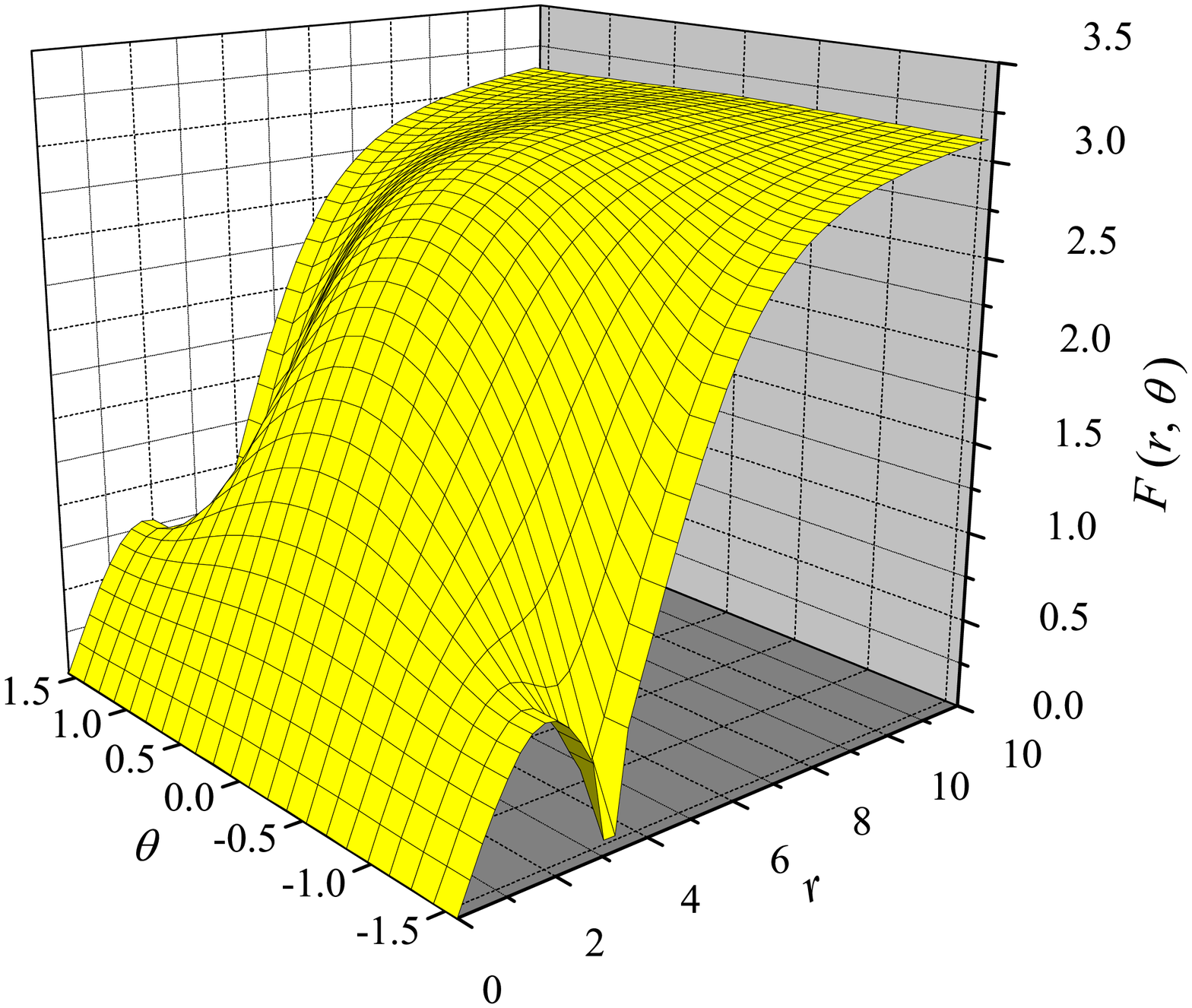}\hspace{-1cm}
\includegraphics[clip,width=80mm]{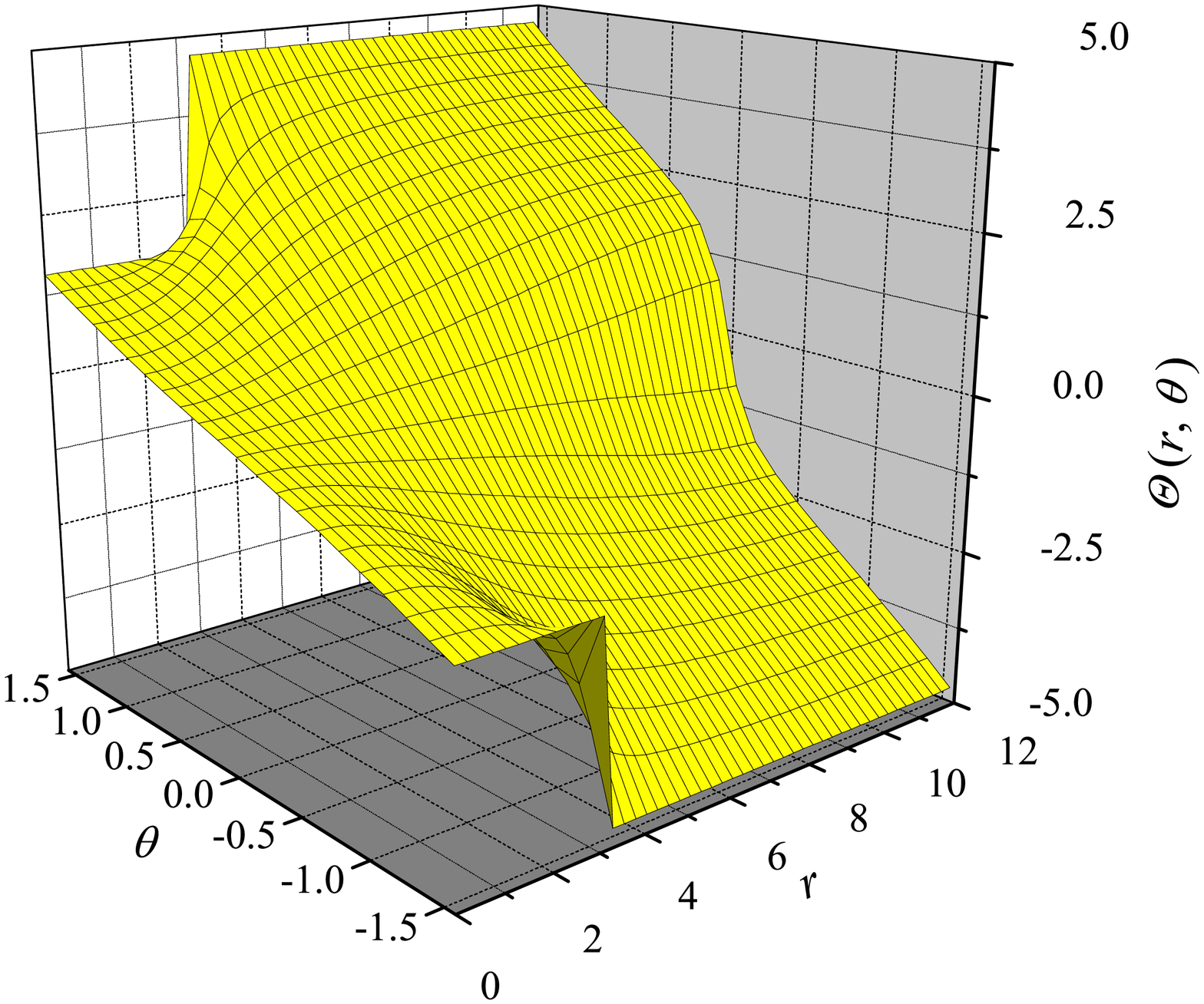}
\caption{\label{fig:profiles_gravity}The profiles of the functions $F(r,\theta)$ (left), $\Theta(r,\theta)$ (right)
 with $\mu=0.08,\alpha=0.005,\beta=0.00,\gamma=0.00$. }
\end{figure*}

\begin{figure*}
\includegraphics[clip,width=80mm]{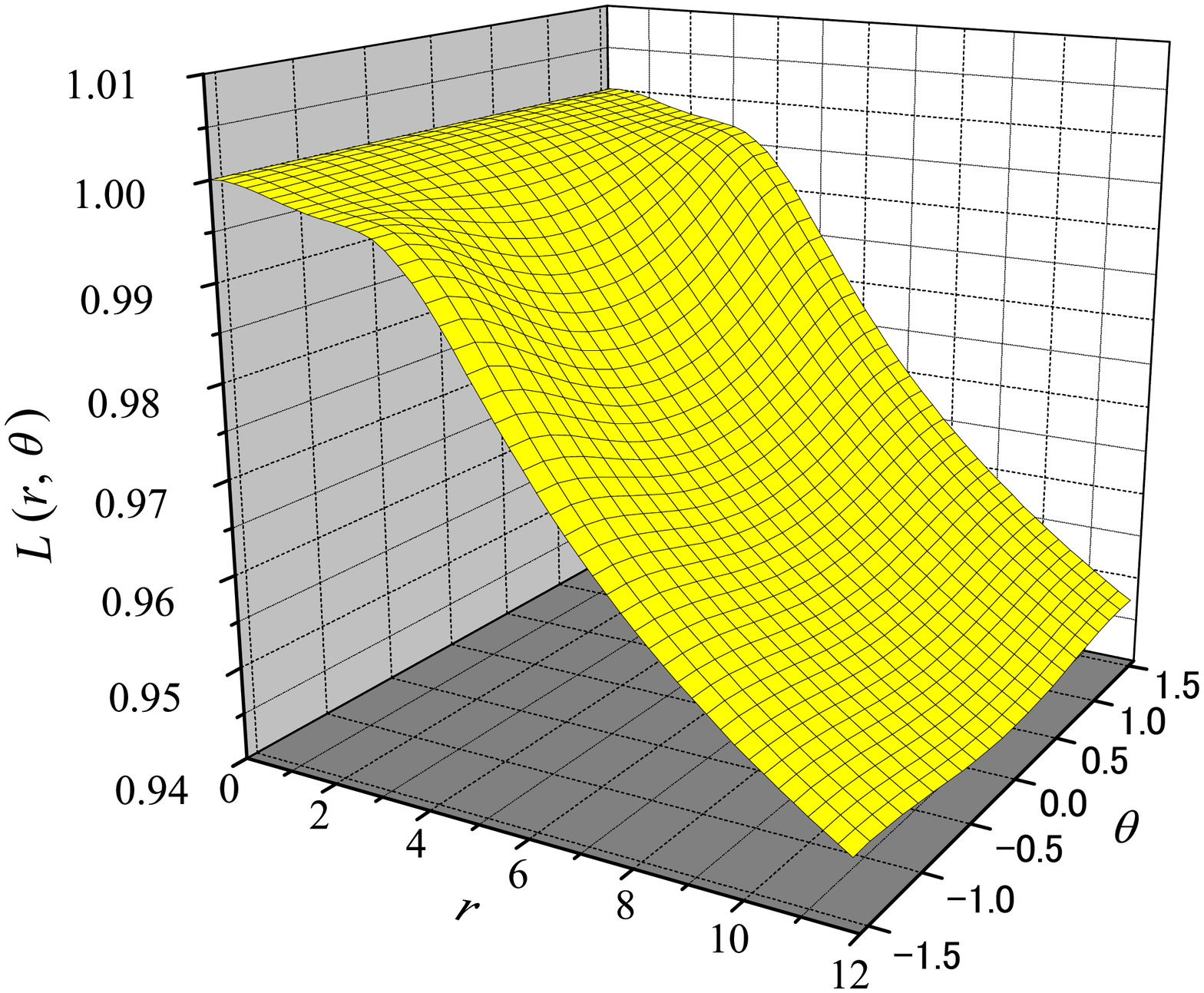}\hspace{-1cm}
\includegraphics[clip,width=80mm]{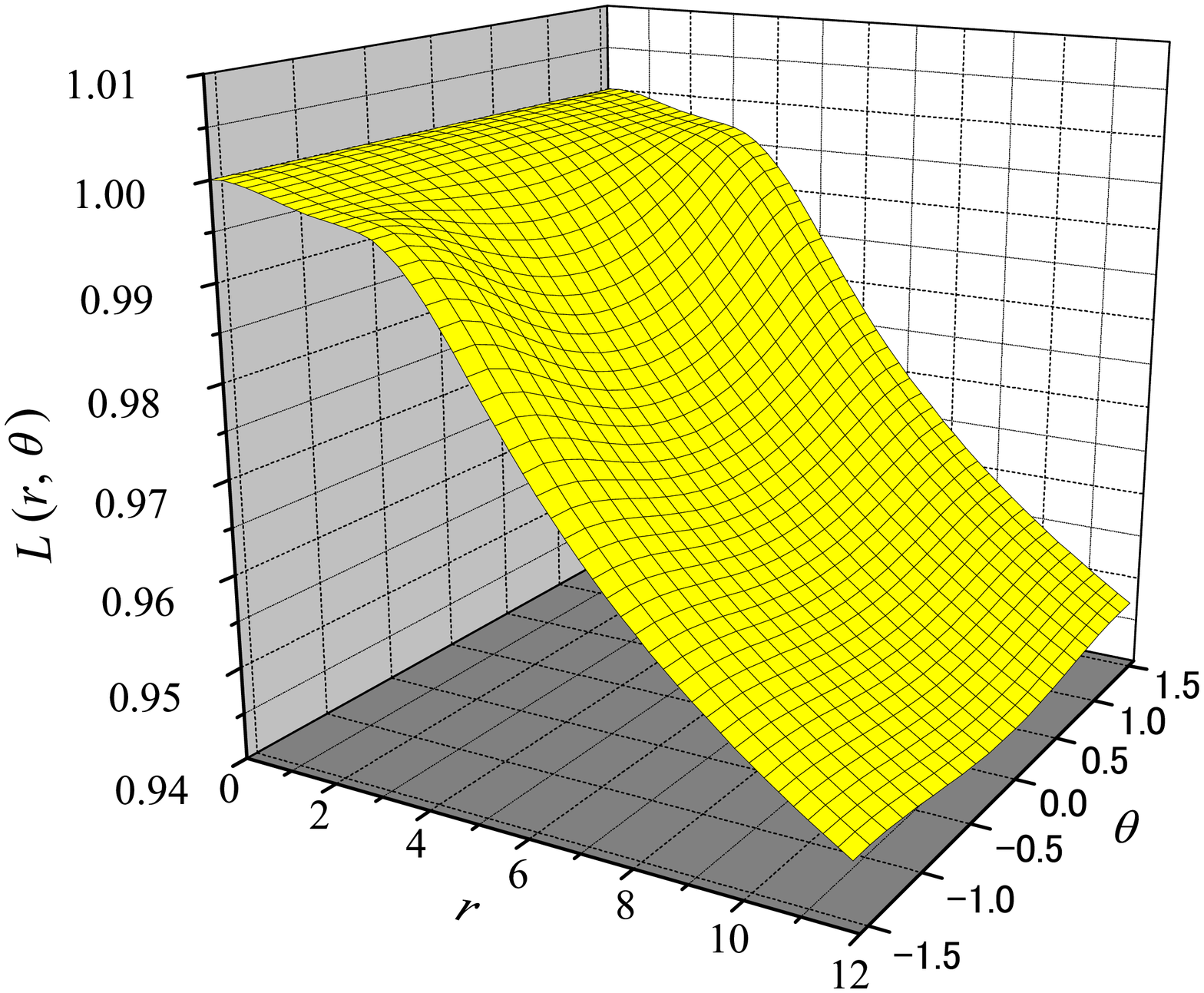}
\includegraphics[clip,width=80mm]{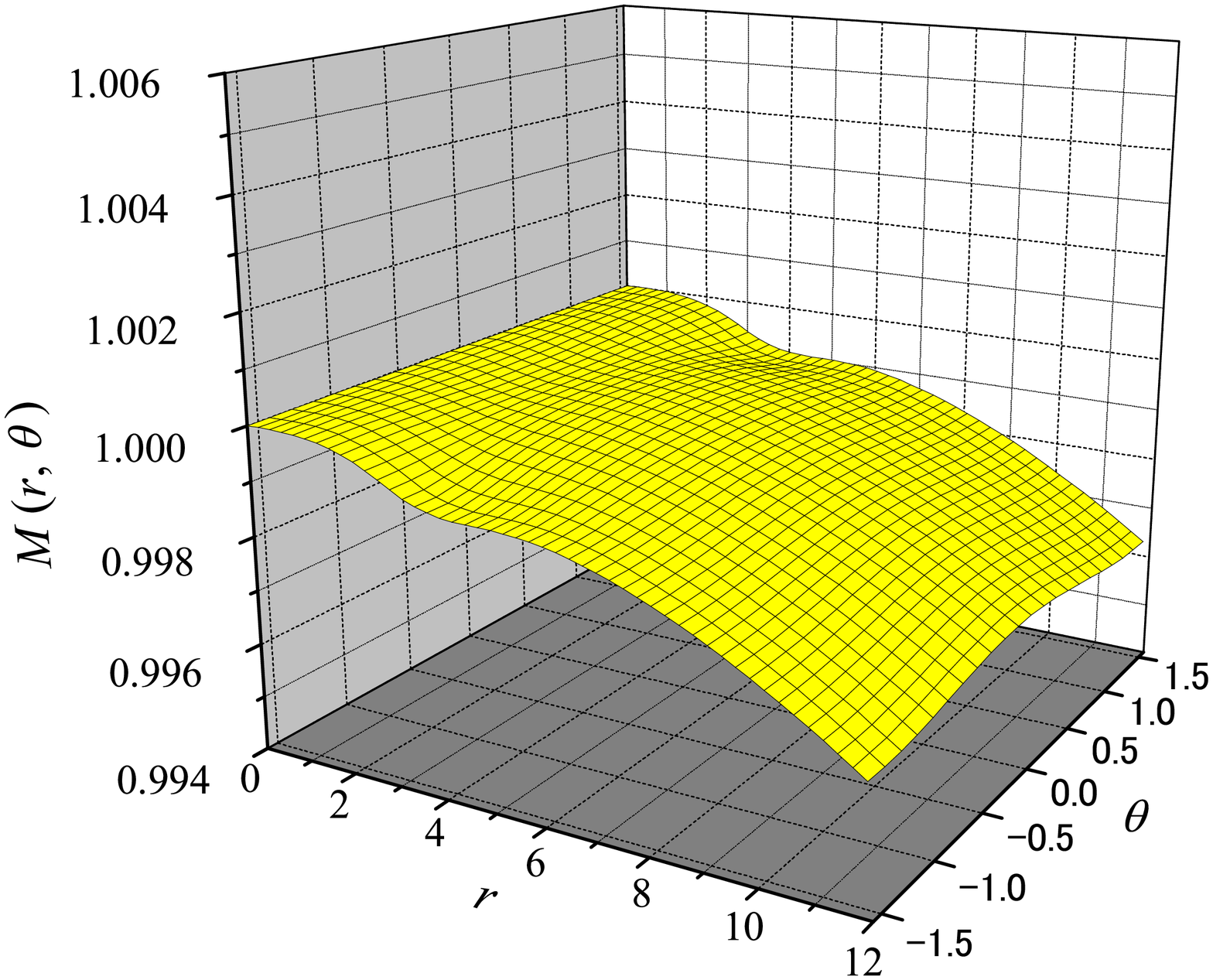}\hspace{-1cm}
\includegraphics[clip,width=80mm]{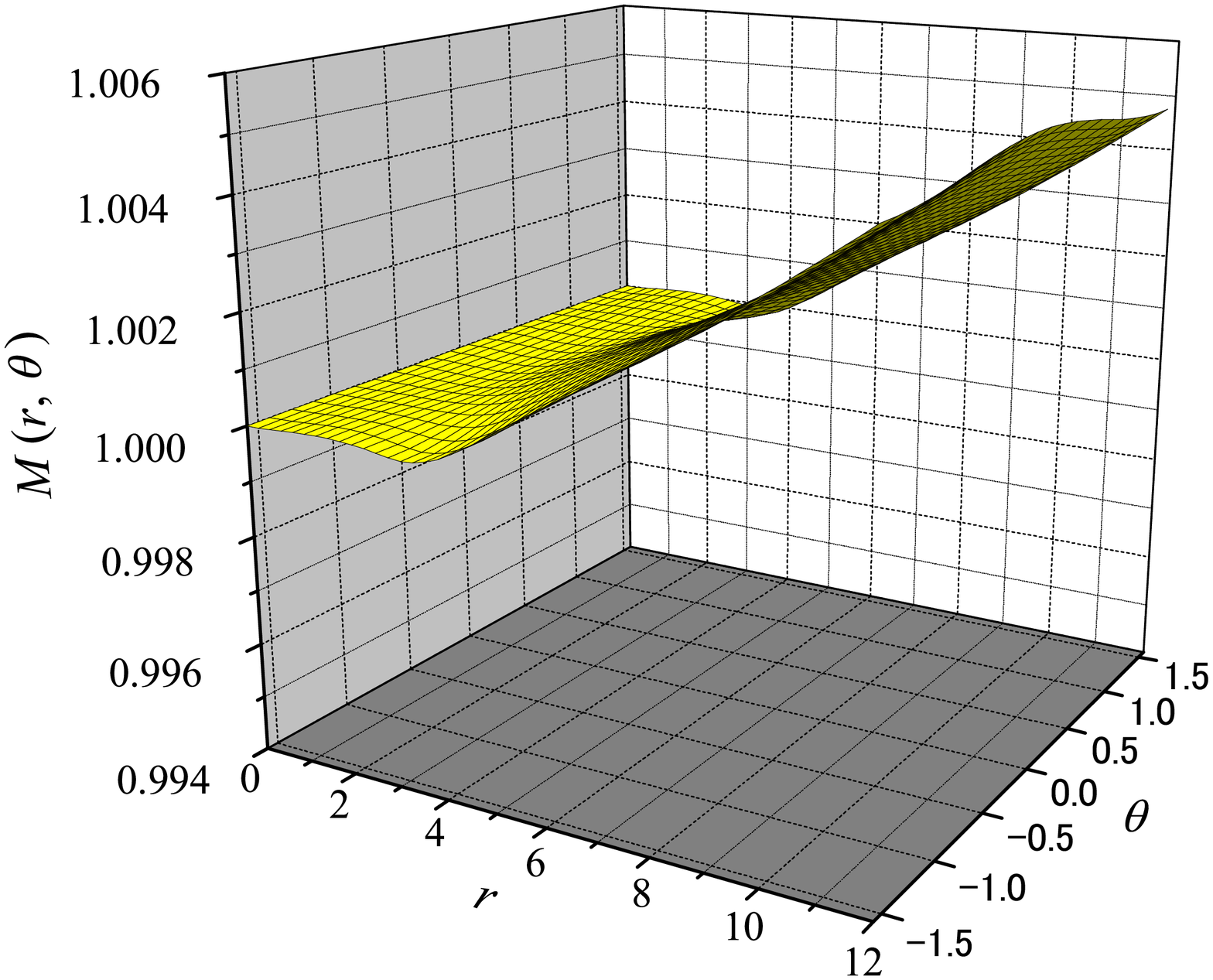}
\caption{\label{fig:metrics}The metrics $L(r,\theta), M(r,\theta)$
for $\beta=0.05$ (left) and $\beta=-0.05$ (right). The remaining parameters are
set to $\mu=0.08,\alpha=0.005,\gamma=0.00$.
}

\end{figure*}

A useful quantity for the gravity-side is the rescaled Ricci scalar which is given by
\begin{eqnarray}
R(r,\theta)&=&
\frac{M^2}{L^2}\Bigl[6\frac{\partial_r^2M}{M}+14\Bigl(\frac{\partial_rM}{M}\Bigr)^2
+\frac{6}{r}\frac{\partial_rM}{M}+2\frac{\partial_r^2L}{L}
-2\Bigl(\frac{\partial_rL}{L}\Bigr)^2+\frac{2}{r}\frac{\partial_rL}{L}\Bigr] \nonumber \\
&+&
\frac{M^2}{r^2L^2}\Bigl[6\frac{\partial_\theta^2M}{M}+14\Bigl(\frac{\partial_\theta M}{M}\Bigr)^2
+2\frac{\partial_\theta^2L}{L}
-2\Bigl(\frac{\partial_\theta L}{L}\Bigr)^2\Bigr]
-\frac{12\gamma}{M^2}\,.
\label{eq:ricci}
\end{eqnarray}

\subsection{Boundary conditions}
For the $\mathbb{Z}_2$-symmetry, the Einstein equations (\ref{eq:Einstein-eq_t})-(\ref{emtensor}),
and the matter field equations (\ref{eq:equationf}),(\ref{eq:equationt}) must be invariant under the transformations 
\begin{eqnarray}
\theta \leftrightarrow \theta+\pi:~~
F(r,\theta)\rightarrow F(r,\theta+\pi),~~
\Theta_0 (r,\theta)\rightarrow \Theta_0(r,\theta+\pi),~~
M(r,\theta)\leftrightarrow M(r,\theta+\pi),~~
L(r,\theta)\leftrightarrow L(r,\theta+\pi)~~
\label{eq:symmetrym}
\end{eqnarray}
and when we set $\theta=0$ as the axis of symmetry, the equations mast be invariant under the transformations
\begin{eqnarray}
\theta \leftrightarrow -\theta:~~
F(r,\theta)\rightarrow F(r,-\theta),~~
\Theta_0 (r,\theta)\rightarrow -\Theta_0(r,-\theta),~~
M(r,\theta)\leftrightarrow M(r,-\theta),~~
L(r,\theta)\leftrightarrow L(r,-\theta).
\end{eqnarray}
Thus as in the case of non gravitation, it is sufficient to restrict the integration domain to the half-plane defined by
$0 \leqq r \leqq \infty$ and $-\frac{\pi}{2} \leqq \theta\leqq \frac{\pi}{2}$. 
For the matter fields, we employ the boundary conditions 
(\ref{eq:boundarycondition00-r}) and (\ref{eq:boundarycondition01-r1})-(\ref{eq:boundarycondition01-t}). 

For the metric fields, regularity at the core of the brane leads to the following conditions
\begin{eqnarray}
\hspace{0cm}M(0,\theta)=1~~\partial_r M(r,\theta)|_{r=0}=0;~~
L(0,\theta)=1,~~\partial_r L(r,\theta)|_{r=0}=0,~~
{\rm for}~~-\frac{\pi}{2}\leqq \theta\leqq\frac{\pi}{2}.
\label{eq:boundarym_a}
\end{eqnarray}
Thus it is likely that the solutions satisfy the following additional boundary conditions
\begin{eqnarray}
\partial_\theta M(r,\theta)|_{\theta=\pm\frac{\pi}{2}}=0,~~
\partial_\theta L(r,\theta)|_{\theta=\pm\frac{\pi}{2}}=0,~~
{\rm for}~~0\leqq r\leqq \infty.
\label{eq:boundarym_b}
\end{eqnarray}
Note that we do not need to impose any conditions for the metrics about $r_0$
because the local spatial structure of the brane is described by only the matter fields.

\subsection{The potential}
As we already mentioned, in \cite{Hen:2007in} the authors have introduced a two parameters class of potential
\begin{eqnarray}
V=\mu(1+n_3)^s~~~~0\le s\le 4.
\label{eq:potential}
\end{eqnarray}
In the original approach, the authors used dimensionful parameters (which is essentially our $\kappa_4/\kappa_2$) 
denoted by $\kappa^2$ and minimized the energy for changing values of $s,\kappa^2$ with fixed $\mu$. 
The main difference in our analysis is that we vary the parameters of the potential, i.e. $\mu$ and 
successively solve  the equations instead of minimizing the energy.

\section{Numerical results}
In \cite{Brihaye:2010nf}, we have done the analysis for the same model but have assumed axisymmetry from the start. In this case the problem reduces to a system of coupled ordinary differential equations.  
In the present case we have to treat the four coupled partial differential equations (\ref{eq:Einstein-eq_t})-(\ref{eq:equationt}) 
with the boundary conditions (\ref{eq:boundarycondition00-r}),(\ref{eq:boundarycondition01-r1})-(\ref{eq:boundarycondition01-r2}) by using the SOR method.

We use the following form 
\begin{eqnarray}
&&F(r,\theta)|_{\rm initial}:=\pi (1-e^{-Ar^{3}}),~~~~\Theta(r,\theta)|_{\rm initial}:=3\theta \nonumber \\
&&\hspace{1cm}M(r,\theta)|_{\rm initial}=L(r,\theta)|_{\rm initial}=1
\label{eq:initialprofile} 
\end{eqnarray}
as the initial profile of the relaxation scheme. 

Note that the conditions we use, together with the initial profile (\ref{eq:initialprofile}) leads to axisymmetric solution. However, when the deformation 
is relatively small, they are approximatively consistent with the boundary conditions of the non-axisymmetric case. Thus, 
we begin with $r_0=0$ and find the solution. By using this result as a initial profile, the computation for the small finite $r_0$ is performed. 
We repeat the procedure until we get the solutions for sufficient large value of $r_0$.
For the continuous change of $r_0$, we can find minimum of the energy per topological charge. 
Since the true solution should exhibits the minimum energy per charge,  
we employ the $r_0$ corresponding to the minimum energy per charge for the boundary conditions. 

 Since the boundary conditions for the metrics are imposed only at the origin, 
the Einstein equations essentially are 
the initial value problem. Usually it is not straightforward to adapt the 
SOR to the problem. Thus we employ the following procedure. We put an initial guess of the form of $L,M$ at the infinity 
and solve the equations for that condition. 
If the condition is inappropriate, the solution surely exhibits the non-trivial singularity at some location, then 
we explore the proper boundary condition until such singular behavior disappears. 
 
\begin{figure*}
\includegraphics[width=8.0cm]{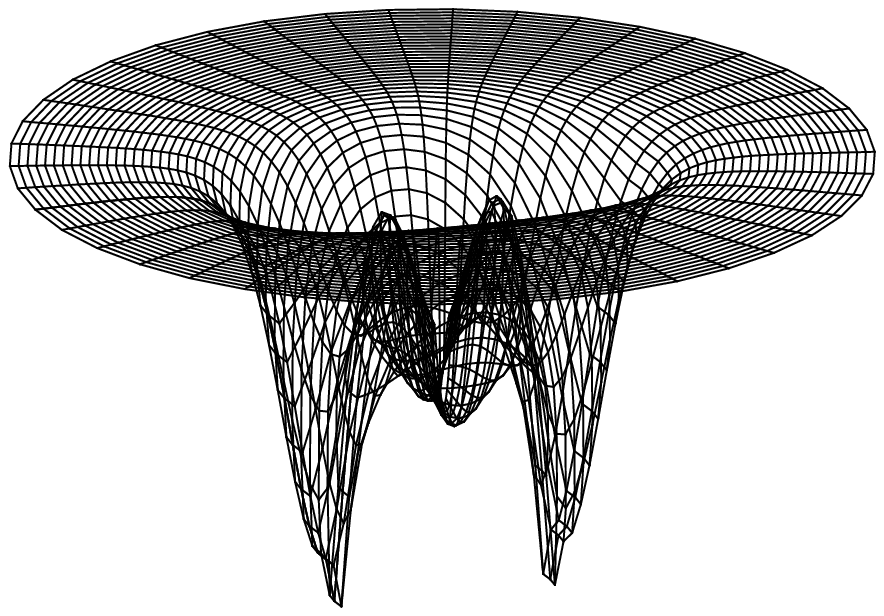}\hspace{-1.5cm}
\includegraphics[width=8.0cm]{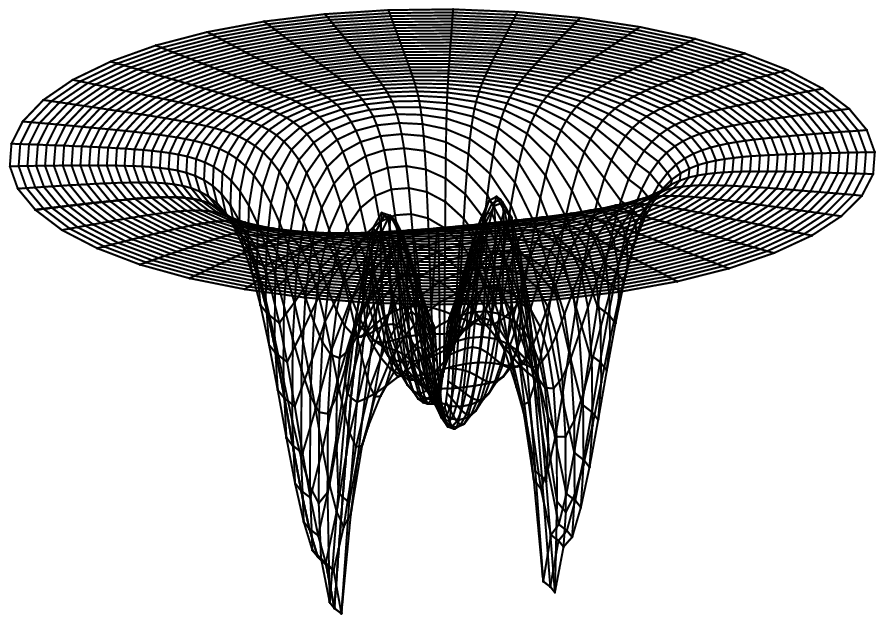}\\
\vspace{-3mm}

{\footnotesize $\beta=0.05$\hspace{5cm}$\beta=-0.05$} \\
\caption{\label{fig:curvature}The Ricci scalar curvature of the $B=3$ solutions for $\alpha=0.005,\gamma=0.0$ .}
\end{figure*}

In Fig.\ref{fig:profiles_gravity} we show a solution for $F,\Theta$.  The metric functions $L,M$ are shown 
in Fig.\ref{fig:metrics} for both sign of the  cosmological constant $\beta$. 
We plot the scalar curvature corresponding to the solutions in Fig.\ref{fig:curvature}. In both cases, there are two
distinct peaks together with the remaining slight ridge. The result for the $\mu=0.08$; if we choose larger value of 
$\mu$, the peaks grow and finally end up in three independent peaks corresponding to the energy density of the baby-skyrmion. 

\begin{figure*}[t]
\includegraphics[width=8.0cm]{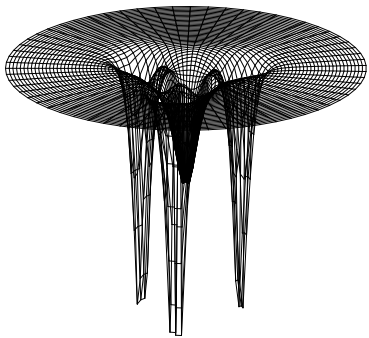}~~~~~~
\vspace{-3mm}

\caption{\label{fig:curvature4}The Ricci scalar curvature of the $B=4$ for $\alpha=0.005,\beta=0.0,\mu=0.01,s=2$.}
\end{figure*}

The Ricci scalar for a typical solution with the $\mathbb{Z}_3$ symmetry is presented in Fig.\ref{fig:curvature4}. 
We clearly see three distinct peaks and for larger value of $\mu$, the fourth peak grows at the origin until we finally get a four-centered 
solution.

\section{Summary}
In this paper, we have obtained the non-axisymmetric brane solutions in the $B=3,4$ baby-Skyrme model with the old-baby potential.
We treated the problem in two steps: we first computed the solution using the SA in order to gain intuition on the boundary conditions. 
Then we solved the four coupled partial differential equations by using the standard SOR by using suitable boundary conditions. 
The axial symmetry breaks to $\mathbb{Z}_2$ symmetry and 
the deformation is a one-parameter family of the strength of the potential $\mu$.

Recently the authors have found at a related model with the present study that there are holomorphic solutions 
for special choice of the potentials~\cite{Ferreira:2011mz,sawado12}. 
The potentials are a class of old-baby potential for $B=1$ while for all other $B$ they become a class of new-baby. 
Furthermore, the axisymmetry is a good symmetry for the new-baby potentials while 
for the old-baby the symmetry is inevitably broken.
The most crucial thing for the deformation is thus whether the value of the potential at the origin is finite or not.  

Study of the linear stability of the present model has been done in \cite{Brihaye:2010nf} within the 
assumption of the axial symmetry. Apart from the axial symmetry, we also need to take into account the 
fluctuation of $\Theta$. It seems straightforward but surely cumbersome. For the level of the classical
solution, our solution is stable for the variation of the size moduli (see Fig.\ref{fig:energy_r0}), 
so we expect that our solution will be stable against any general perturbation as well. 

Of course there are several possibilities of the potential for obtaining the anisotropic solutions. For example, 
in \cite{Jaykka:2011ic} the authors introduced a  more sophisticated potential $V=|1-(n_1+in_2)^N|^2(1-n_3)$~(with the integer $N\geqq 2$). 
However the potential is essentially of old-type and the solution tends to split the number of $B$ constituents. 
An interesting feature of such models is the fine structure: the zero of the potential are at the infinity $(F=0)$ as well as the
points $F=\frac{\pi}{2}$ and $\Theta=\frac{n\pi}{N}$ ($n$:integer). As a result, the solution
split $B$ component where each has $NB$ peaks. 

An intersting question raised by this work is to see the effects of the the spatial structure of the branes on the 
mass of the brane localized fermions (quarks, leptons). 
The fermions coupled with the baby-skyrmions with Yukawa-coupling exhibit non-trivial doubly degenerated 
plus isolated states which could be identified as three generations of the quarks/leptons. The model has 
weak isospin symmetry and the isodoublet of the quarks are degenerate.  
We expect that the brane slightly deforms from the axisymmetry and reproduce the slight 
splitting the mass of the first and second levels~\cite{Kodama:2008xm}. Now we got the intuition that   
the level splitting of the generations is owed by the coupling constant of the old-baby potential $\mu$. 
More thorough analysis of the property of the localized fermions 
in these funny structures of the brane will be certainly very interesting.  

\appendix
\section{Topological charge}
The topological charge defined by (\ref{topologicalcharge}) is easily computed for the boundary conditions 
(\ref{eq:boundarycondition00-r}), (\ref{eq:boundarycondition01-r1})-(\ref{eq:boundarycondition01-t}). From (\ref{topologicalcharge_polar})
\begin{eqnarray}
B&=&-\frac{1}{2\pi}\int_{-\frac{\pi}{2}}^{\frac{\pi}{2}}d\theta\int_{0}^{\infty}dr~
\bigl[\partial_r (\cos F\partial_\theta \Theta)-\partial_\theta (\cos F\partial_r \Theta)\bigr] \nonumber \\
&=&-\frac{1}{2\pi}\int_{-\frac{\pi}{2}}^{\frac{\pi}{2}}d\theta~
\bigl[\cos F\partial_\theta \Theta\bigr]_{0}^{\infty}
+\frac{1}{2\pi}\int_{0}^{\infty}dr~
\bigl[\cos F\partial_r \Theta\bigr]_{-\frac{\pi}{2}}^{\frac{\pi}{2}}\nonumber \\
&=&-\frac{1}{2\pi}\int_{-\frac{\pi}{2}}^{\frac{\pi}{2}}d\theta~
\bigl(\cos F|_{r=\infty}\partial_{\theta}\Theta|_{r=\infty}-\cos F|_{r=0}\partial_{\theta}\Theta|_{r=0}\bigr) 
\nonumber \\
&&+\frac{1}{2\pi}\int_{0}^{\infty}dr~
\bigl(\cos F|_{\theta=\frac{\pi}{2}}\partial_{r}\Theta|_{\theta=\frac{\pi}{2}}-\cos F|_{\theta=-\frac{\pi}{2}}\partial_{r}\Theta|_{\theta=-\frac{\pi}{2}}\bigr)
\nonumber \\
&=&-\frac{1}{2\pi}\bigl((-1)\cdot3\pi-1\cdot\pi\bigr)+\frac{1}{2\pi}\bigl(1\cdot\pi-1\cdot(-\pi)\bigr) \nonumber \\
\nonumber \\
&=&3
\end{eqnarray}
where we used the relation 
$\pm \frac{1}{\pi}\partial_r\Theta |_{\theta=\pm\frac{\pi}{2}}=\delta (r-r_0)$.

\end{document}